\documentclass[aps,amsmath,prb,a4paper,floatfix,twocolumn]{revtex4}
\usepackage{graphicx}
\usepackage{subfigure}
\usepackage{stackengine}
\usepackage{amsmath}
\usepackage{amsfonts}
\usepackage{amssymb}
\usepackage{mathtools}
	\usepackage[latin1]{inputenc}
	\usepackage[T1]{fontenc} 
	\usepackage{fancyhdr}
	\usepackage{dsfont}
	\usepackage{setspace}
	\usepackage[italian]{varioref}
	\usepackage[colorlinks=true,linkcolor=black,urlcolor=black,citecolor=black]{hyperref}
	\usepackage{makeidx}

\newcommand{\beq}{\begin{equation}}
\newcommand{\eneq}{\end{equation}}
\newcommand{\bea}{\begin{eqnarray}}
\newcommand{\enea}{\end{eqnarray}}
\newcommand{\met}{\frac{1}{2}}

\newcommand{\RR}{\mathbb{R}}
\newcommand{\wt}{\widetilde}

\usepackage{xr}

\begin{document}



\vspace{5cm}

 \begin{figure}[h]
\begin{center}
\includegraphics[scale=0.6]{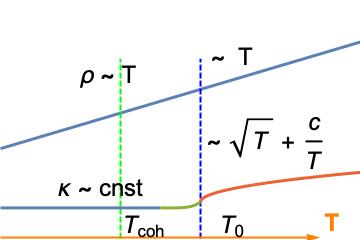}
\caption{sketch of the temperature dependence of  the thermal conductivity $\kappa$ and of the resistivity $\rho$  in the incoherent  phase of the strongly interacting fermionic  model presented here, which is a sort of  extension to higher  space dimensions of the $0+1-d$  SYK model in its  pre-chaotic phase. } 
\label{tritoc}
\end{center}
\end{figure}

\hspace{8cm}\\

\hspace{8cm}\\
\hspace{8cm}\\
\hspace{8cm}\\

The $0+1-d$  Sachdev-Ye-Kitaev (SYK) fermionic model  attracts  nowadays a wide spread  interest of the  Condensed Matter community, as a benchmark  toy model for strong electron correlation and non Fermi Liquid behavior.  It is exactly solvable in the  infrared limit and reproduces  the linear dependence  of the resistivity on temperature $T$, in linear response, typical of the strange metal phase of High Temperature Superconducting (HTS)  materials. The breaking  of its conformal symmetry requires ultraviolet corrections for a  faithful description of its pseudo Goldstone Modes. Extension of  the model  to higher space dimension includes a local U(1) phase generating collective bosonic excitations driven by the additional ultraviolet contribution to the action.   These  excitations  are studied here, in a temperature window of   incoherent dynamics   in which the expected chaotic regime has not yet taken over.  We identify them as   neutral  diffusive energy excitations with temperature dependent  lifetime $\sim \hbar / k_B T$. They provide   thermalization of the system  and contribute to  the $T$ dependence of the transport coefficients. The linear  unbound   $T$  increase  of particle current is confirmed by our  hydrodynamic modelization. A  quantum liquid in interaction with this system would become a Marginal Fermi Liquid.

\title{ The extended diffusive Sachdev-Ye-Kitaev model as a sort of "strange metal"}
\author{  A.Tagliacozzo$^{1,2,3}$}
\affiliation{$^{1}$ INFN-Sezione di Napoli, Complesso Universitario di Monte S. Angelo Edificio 6, via Cintia, I-80126
Napoli, Italy}
\affiliation{$^{2}$ Dipartimento di Fisica "E. Pancini", Universit\`{a} degli Studi di Napoli Federico II, via Cintia, I-80126 Napoli, Italy}
\affiliation{$^{3}$ CNR-SPIN, Monte S.Angelo via Cintia, I-80126 Napoli, Italy}

\maketitle


\vspace{0.5cm}

\section{Introduction}
After almost three decades the strange  high temperature metal phase of the various families of materials which become  superconducting  at a high critical temperature $T_c$   and are denoted as High Tc Superconductors (HTS) is still a puzzle. The great number of families of materials which undergo superconductivity at high temperature, cuprates, iron based material and picnides, possibly nickelates,  present   rather general and basic  surprising features, which  require  an advancement in the understanding of  strongly correlated fermion systems.  The most striking of these features  is the linear increase with temperature of the electrical conductivity in the normal metal phase, up to $\sim  1000\:  K $, without any sign of saturation\cite{gurvitch,daou,chaWentzell} and the constancy of the thermal conductivity\cite{cohn}. The second puzzling feature is the misterious mechanism for superconducting electron pairing at "high" temperature.

The phenomenon of superconductivity has always been an icebreaker vessel for the interpretation  of the quantum  nature of the  metal realm.  The BCS  quantum model came out more than fourty years  after the discovery of superconductivity in metals, most likely because  the full fledge Quantum Field Theory was required to provide a  convincing tool box. 
Something similar is happening today with  the HTS, because we do not have full understanding of strong correlation in many body systems. 
 It was immediately realized by P.W.Anderson\cite{anderson} that the doped Mott insulator, which  appears as a good  starting point for interpreting  the  normal metal phase of HTS, breaks the frame of the Landau Fermi Liquid (FL) theory\cite{hill,parc},  a theory which  has dominated  successfully  up to the eighties. A brand new approach to strong  correlation in metals, the Non Fermi Liquid Theory (NFL)\cite{schofield,varmaNussinov}, is required,  which is still unsettled. 
 
A  burst of interest comes now  from the appearance of a new  exactly solvable model in $0+1-d$ dimension, the Sachdev-Ye-Kitaev (SYK)\cite{sachdevye,kitaev,kitaev06} model,  which describes  random all-to-all  ${\cal{J}}-$ interaction between $N$ Majorana fermions. The SYK model  has become highly popular as a holographic dual for gravity theories of black holes \cite{maldacena98,sach10,policastro}. Although it appears as the farthest possible from a  realistic description of any real material,  its extension to higher space dimensions could make it  a useful prototype model for NFL.

   Generalized SYK models have been proposed with extension to higher space dimensions \cite{lantagne,guQi,davison,song,berkooz,chew,haldarBanerjee,patel18B,chowdhurySenthil}, most of them  having in mind applications to High Critical Temperature ($HT_c$) superconducting materials. There are complex fermion versions of the SYK model\cite{fuSachdev,chaWentzell,davison,klebanov}. 

    Lattice anisotropy which characterises the cuprates and their  $2-d$ $CuO$ planes would motivate preference for  two space dimensions.  We are unable to bridge "bad" or "strange" metals and Quantum Spin Liquids on one side\cite{varma,carlson,jain}  with the SYK  model on the other side\cite{sach10,patel,haldarShenoy,chowdhury20A, chowdhury20B,davison,cha}, but we concentrate on  the circumstance of approximate symmetry breaking in the infrared (IR) limit, when there is  scaling to strong interaction $ {\cal{J}} \to \infty, N\to \infty$, but  finite  $\beta {\cal{J}}/N $ ratio.  Fermionic  pseudo Goldstone modes (pGm)  arise and  additional  bosonic excitations,  when  the hopping on a  $2-d$  space  lattice also  includes  the first UV corrections\cite{noi}.   
    
      In fact, in the IR  limit, hopping  in the lattice can be dealt with within lowest perturbative order.  The response of the fermionic excitations,  in the conformal symmetry limit, to  an external driving to be specified,  gives rise to the celebrated linear temperature dependence of the resistivity and to the constancy in temperature of the thermal conductivity.   However, the kinetic term added to the model  requires the addition of a complex   $U(1)$ phase to the real fields, which couples with  the  fermionic pGm  beyond the IR limit and gives rise to bosonic  collective and diffusive modes\cite{song,cai18A,noi}.  The original part of this work focuses on these  modes, which we call Q-excitations. We set up an hydrodynamical description of these  energy modes which, by  transporting  energy,  are responsible for the thermalization of the system in a temperature window $ T_{coh}\lesssim T \lesssim T_0$.  The temperature  $T_0$, with  $ \beta T_0 \sim {\cal{O}} \left (\beta {\cal{J}}/N \right )$  (the notation $ {\cal{O}} ( ...)  $ means  "order of "), is given by Eq.(\ref{tdisst}) and  is assumed to be  the temperature threshold above which, on the basis of what found in the $0+1-d$ SYK model,  a "scrambled phase" leading to quantum chaos is expected  in the SYK system.  
      
      On the other side, $T_{coh}$ denotes  the temperature  below which hopping in the lattice is coherent. If $W$ is the bandwidth which arises from the hopping in the lattice,  $ T_{coh} \sim W^2/ N {\cal{J}}$, as argued in Section II.  To guarantee that hopping is a marginal perturbation in the  $ {\cal{J}}, N\to \infty$ limits,  is $ \beta T_{coh}\sim {\cal{O}} \left (\beta {\cal{J}}/N \right )$, as well.
      
          \begin{figure}[h]
\begin{center}
\includegraphics[scale=0.4]{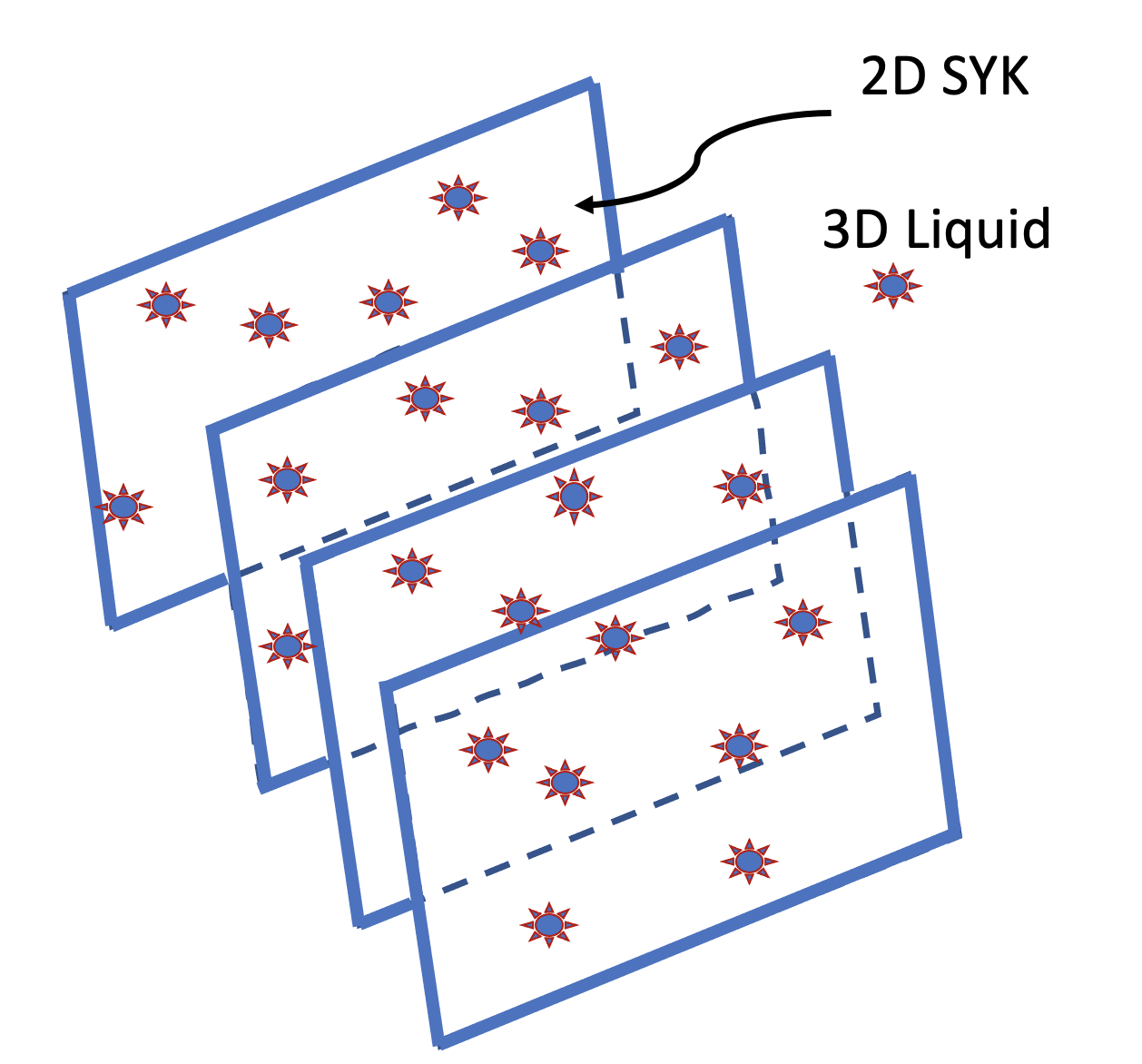}
\caption{A caricature of the toy  system modelized  in this work.  $2-d$ sheets  of strongly interacting Majorana Fermions are described by  a SYK model  which includes a hopping term in the continuum limit approximation. The SYK  action is given by  Eq.(\ref{aca}).  In the temperature window considered here the SYK system is incoherent  and fully thermalized, but  not chaotic.   Linear response of the SYK system in the IR limit  gives a  thermal conductivity  which is   $T$  independent and  a linear in $T$ resistivity. Breaking of the conformal symmetry which is present in the IR limit require  UV corrections  for  regularization.  Bosonic collective  diffusive  Q-excitations  arise,  which contribute to the transport coefficients. A $3-d$  single band   Fermi Liquid (FL)   of  complex fermions  ("electrons" represented as stars in the picture)  becomes a  Marginal Fermi Liquid (MFL), when perturbative interaction  with the SYK sheets in the incoherent  IR limit  is turned on.  Back-interaction between the SYK sheets induced by  the coupling with the MFL  is not considered. The interaction of the  Q-excitations with the MFL  do not  change the linear $T$ dependence of the particle current derived in the IR limit, but adds extra $T$-dependence  to the thermal conductance.  } 
\label{model}
\end{center}
\end{figure}

       Section II  collects the  properties of the   $0+1-d$ SYK model perturbed by the lattice,  in the conformal symmetry limit, including the hopping term in the action. Fermionic quasiparticles  can be derived which smear the Fermi Surface (FS), but the most relevant contribution to particle and thermal conductivities comes from the dominant pGm which arise from the approximate symmetry breaking. These incoherent excitations are the sources of  space gradients  of the chemical potential $\mu$, which is related to the  time derivative of the phase $ \partial _t \varphi$, introduced by  the hopping term in the extended action. We characterize these phase modes and the related response of the system with the help of  a hydrodynamical  dissipative approach. The linearity of the  $2-d$ resistance $\sigma^{-1}$   with temperature and the constancy of the thermal conductivity $\kappa$  are obtained in this way, as reported in Section II.C. They are  ${\cal{O}} \left (\beta {\cal{J}}/N \right )$  in the large $N$ strong ${\cal{J}}$limit. 
       
       From now on we investigate the single particle and collective excitations of the model as well as their interactions to  give an estimate of further contributions to the transport coefficient arising from these extra excitations beyond the IR  limit.  The bottom line of this analysis  is that all extra contributions to $\kappa$ and $\sigma^{-1}$ coming from these extra excitations are are subdominant with respect to the ones  of Section II.C and can be ignored in the strong coupling limit.  From Section III onward, corrections beyond the IR approximation are considered.
       
      The phase $\varphi$   arises in a reparametrization of the fermionic propagator in the IR  limit.  This allows to extract the first correction to the action  $  \sim {\cal{O}} \left (N/\beta {\cal{J}}  \right )$, i.e. the one coming from ultraviolet (UV)  corrections. This extra term entails a diffusive dynamics  for the phase and its fluctuations. In the absence of  coherence, hydrodynamics  can help to extract the collective properties of the phase  beyond the conformal limit, as  discussed in   Section III.  It is assumed that thermalization occurs at a Planck rate, with lifetime of the collective excitations  $\tau_0 \sim \hbar \beta$ (where  $\beta = 1/k_BT$).  A  temperature independent  diffusion coefficient $ \tilde{D}_Q$ can be derived  in this frame, which turns out to be ${\cal{O}} \left (\beta {\cal{J}}/N \right )$,  and allows for the  definition of  a mean square  diffusion length $\tilde{a}_\ell \sim T_0/T$.  In this approximation, the gapless diffusive Q-excitation modes  contribute to the thermal conductivity with a term  $\propto \sqrt{T}$, which  is subdominant, being  ${\cal{O}} \left ([\beta {\cal{J}}/N ]^{1/2}\right )$ (Section III.B).

      In Section IV  we consider  an hypotetical  $3-d$ quantum Liquid (qL), in which   SYK sheets are   embedded,  which would be a Fermi Liquid of the same  bandwidth $W$, when considered in isolation(see Figure \ref{model}). The complex fermionic quasiparticles  of the  qL, close to the Fermi energy become ill defined when the local interaction with the $0+1-d$ SYK  systems is turned on  the qL becomes a Marginal Fermi Liquid (MFL).  Inserting the quasiparticle lifetime of the MFL in an approximate expression for the thermal conductance, Eq.(\ref{ktMFL}), provides again a subdominant contribution to $\kappa$, which is constant with temperature. 
 In Section IV.B,  the additional scattering rate induced by the presence of the Q-excitations is derived by means of the Fermi Golden rule,  and compared to the case of the usual electron-phonon interaction.  Competing effects produce a big drop of the exponent in the  $T-$power law, with respect to the electron-phonon case. On the one hand  incoherent scattering,  in which thermalization is very effective, so that no bookkeeping for particle number and momentum is maintained,  should imply higher scattering possibilities and higher scattering rates. On the  other hand, the  very local nature of the interaction reduces the energy matrix element and the scattering rate. The quasiparticle inverse lifetime  acquires a FL-like contribution $\sim T^2$ to be summed to the MFL scattering rate in a sort of  Matthiessen's rule. This $ T^2$ inverse scattering rate implies a $1/T$ contribution to thermal conductivity which is however subdominant once more. Similarly for the  resistivity contribution, which, though subdominant,   turns out to maintain the experimentally measured linear $ T$ dependence\cite{gurvitch}. 
 
  In the Appendix A, the  derivation of the hydrodynamic relativistic equations used in the text is reported.

 \section{the SYK model and its extension to higher space dimensions}
 The   Hamiltonian $ {\cal{H}}_0 $ for the uncoupled  $0+1$-d SYK dots  is:
    \bea
    {\cal{H}}_0 =  \sum_x  H_x =   \frac{1}{4!}\:  \sum _x \sum _ {klmn} {\cal{J}}_ {klmn}    \chi_{x,k}  \chi_{x,l}  \chi_{x,m}  \chi_{x,n}.
     \label{je0}
     \enea
      Lattice position is  denoted by the subscript $x$ in the continuum limit.  $ \chi_{x,l} $ are Majorana fermion operators of $N$ fermionic flavors  ($ {klmn} \in 1,..,N$) on  each site $x$ . 
      
      In each uncoupled dot,   the all to all interaction  $ {\cal{J}}_ {klmn}  $  is  averaged over.  After this averaging, the incoherent dynamics  produced by  the random interaction and disorder  is parametrized by an average  interaction strength ${\cal{J}}$.  The large ${\cal{J}}$ and  $N$ limit  (infrared  (IR) limit) is considered, which is  exactly solvable. In the IR limit, the model  develops an approximate conformal symmetry,  because it is  invariant under a full reparametrization  symmetry.   Spontaneous   breaking  of this symmetry  occurs, down to the $\wt {SL}(2, \RR)$ group symmetry\cite{kitaevGeom}, with pseudo-Goldstone modes (pGm)  which make the four point function divergent. The  two point function, the time ordered single particle  Green's function, $G^\beta_c(\tau) = {\bf Tr} \left \{ T_t  \left [  \chi_{x1}(\tau)  \chi_{x1} (0) \right ] \right \}$ plays the role of an order parameter and is non local in imaginary time $\tau$. However,  symmetry breaking is  induced anyhow,  when  the first  ultraviolet correction is included to cure the divergency\cite{maldacena,KitaevSoft}.  An additional  term  of ${\cal{O}}\left ( N/\beta {\cal{J}}\right )$ is added to the action,  as a result of reparametrization of the"quasi-order-parameter" field.  While the latter  is bilocal in time, the added correction to the action is local in time and adds local dynamics which becomes extended  in space when the model is extended to higher space dimensionality. 

 While the Green function and the self-energy of the original model  are real, those of the extended model, 
   $G _x(\tau_1,\tau_2) ,\: \Sigma_x (\tau_1,\tau_2)$, have a   space dependent modulus and  phase, with small fluctuations around   the saddle point   values $G_{c }(\tau_{12}), \: \Sigma_c (\tau_{12}) $  of the $0+1-d$ SYK dot at each  site $x$ (here $\tau_{12}\equiv  \tau_{1}-\tau_{2}$). We write:
 \bea
G_{x}(\vartheta_1,\vartheta_2) = \left [G_{c }(\vartheta_{12})+ \delta G(x,\vartheta_{12}, \vartheta_+ )\right] \: e^{i \: \varphi _x(\vartheta_+) },\nonumber\\
 \Sigma_{x}(\vartheta_1,\vartheta_2) = \left [ \Sigma_{c }(\vartheta_{12})+ \delta \Sigma (x,\vartheta_{12}, \vartheta_+)\right ] \: e^{i \varphi _x(\vartheta_+) }.
 \label{greo}
\enea
Here   $ \vartheta= 2\pi \tau / \beta $ is a dimensionless time and   $\vartheta_{12}=\vartheta_1-\vartheta_2$, while 
   $\vartheta _+ = (\vartheta_1+\vartheta_2)/2$  is a local time in the space evolution   (however,  we will most of the times denote the dimensionless time as $\tau$ unless differently specified).

      The local correction $ \delta G\left (x,\tau_1-\tau_2, \tau_+\right ) \: e^{i \: \varphi _x(\tau_+) }$  arises from    nearest neighbour  hopping   in a lattice of spacing $ {\tilde a}$. 
      A simple argument for the determination of the relevance of the kinetic energy term in the  scaling to the UV is reported here below.  
      
      Let  
 $ \tau ( \equiv \tau_+) \to \tau ' = \tau /s$ with $ s  \to  {\cal{J}} /T$. Majorana fields scale as $  \chi'  \sim \chi \: s^{1/4} $ (scaling dimension is $\Delta = 1/4$). The kinetic energy due to the lattice is naively written as 
 \bea
  t_0 \int d\tau \: \chi_{x,l} \chi_{x',m} \to  t'_0 \int d\tau ' \: {\chi'}_{x,l} {\chi'}_{x',m}, 
  \enea
   with $    t'_0 \sim s^{1/2}  t_0 $.  This  is a relevant perturbation. Scaling up to $ s_T =  {\cal{J}}/T$ implies 
  \bea
 t_0 \to t_0( s_T) = t_0\times  \left ( \frac{  {\cal{J}}}{T} \right ) ^ \met ,
\enea
which is a weak hopping, up to $t_0  ( s_T)  \sim  {\cal{J}}$. This  defines a threshold  ( 'coherence' ) temperature  $ T_{coh} $ when:
\bea
  {\cal{J}} \sim   t_0 \times \left ( \frac{  {\cal{J}}}{T_{coh}} \right ) ^ \met \:\: \to \: T_{coh}  \sim \frac{t_0^2}{ {\cal{J}}} .  
 \label{tcoh}
 \enea

 For $T > >T_{coh}$ the lattice can be dealt with as  an irrelevant  perturbation  on the $0+1-d$ SYK model, while  for $T < T_{coh}$,  the hopping is a marginal correction in the scaling limit. 

       We introduce now a more careful approach to  the kinetic energy term to be added to Eq.(\ref{je0}). 
     
     Electron quasiparticles  hopping  out of   site $x$ to  one of the  neighbouring  sites  requires complex fermions.  In a full paramagnetic system we drop  out the spin,   and  the complex fermionic spinless operator  $  c_x^\dagger , c_x  $  for the particle  can  
   be   represented   in terms of two  flavours  of  the neutral fermions on the same site:
     \bea 
     c_x = \frac{1}{\sqrt{2}} \left ( \chi_{x1} +i \: \chi_{x2} \right ), \:\:\:  c_x^\dagger  = \frac{1}{\sqrt{2}} \left ( \chi_{x1} -i \: \chi_{x2} \right ).
     \label{opert}
     \enea
      The hopping  will  change  the interaction energy of all the Majorana Fermions of  site $x$.  As we  ignore conservation of the particle number, we focus just on the site which is the origin of the jump. However, the local  interference among  neighbouring sites involves  the local phase factor $ e^{i \: \varphi _x(\tau_+) }$. It can be shown\cite{noi} that  the change of energy is, for one single realization of disorder and replica:
       \bea
 {\cal{H}}_{Kx} (\tau) =\! i\: \frac{1}{3!} \sum _{lm} J_{12lm} e^{i \: \varphi _x(\tau_+) } \chi_{x,l}(\tau)  \chi_{x,m}(\tau)  + h.c. 
 \label{hamom}
\enea     
    
  
 The thermal average including annealed disorder (the average uses  the replica trick in  the standard SYK model)   includes now  an additional  gaussian average for $J_{12lm}$, which  mixes terms from different sites, and gives  rise to  random interdot hopping of energy scale $t_0/N$.  As usual,  we have assumed that the mixing of replicas can be ignored in the saddle point approximation and we have dropped the replica label  everywhere.  The temperature determines  the relevance  of the resulting  kinetic term in the limit $ {\cal{J}},N \to \infty$.  To guarantee that the kinetic term is  marginal also in  the   incoherent  phase limit, it is  required that $\beta T_{coh}  \sim {\cal{O}}\left (\frac{ \beta {\cal{J}}}{N} \right )$, where $T_{coh}$ is defined  as  $ \frac{t_0^2}{N{\cal{J}}}$,  as suggested by   Eq.(\ref{tcoh}).

 The next step is  the integration over the Majorana fields  $ \chi_{x,l}(\tau)$,  with the help of the Hubbard-Stratonovich fields  of Eq.(\ref{greo}). The saddle point  action in terms of the complex bilocal auxiliary fields $ G_x( \tau_1, \tau_2) $ and $ \Sigma_x( \tau_1, \tau_2) $, characterized by the  phase $\varphi _x $, is:  
\begin{widetext}
  \bea
  \frac{I_{ex}}{N}  =\sum _x\left [ - \met \ln  det\left [ \partial _\tau \!- \Sigma_x \right  ] \!+ \!\int \!d\tau\: d\tau ' \!\left \{\! - \frac{ J^2 }{4}\left | G_x(\tau,\tau')\right |^4 \!+\! \Sigma_x(\tau,\tau')G_x^*(\tau,\tau') -\frac{t_0^2}{N} \! \sum _{x'\in  nn} G_x(\tau,\tau')G_{x'}^*(\tau,\tau')\!\right \} \right ].
  \label{aca}
  \enea
 \end{widetext}
 The last  term of the action  is the intersite kinetic term, which  was also previously derived\cite{song}.  The action has  become complex due to the  additional $U(1)$ minimal coupling in Eq.(\ref{hamom}).
  
The real  Green function in the saddle point approximation is 
 \bea
  \frac{1}{G(i\omega_n) }  = -i\omega _n- \Sigma ( i\omega_n ), \:\:\:\:  
    \label{galt1}
 \enea
  where  $i\omega_n$ are fermionic frequencies. 
  
  The conformal limit  in the IR corresponds to the dropping  of $\partial _\tau$ in the determinant , or  of $ -i\omega _n$ in Eq.(\ref{galt1}). In this limit,  at  low temperature,  assuming particle-hole (p-h) symmetry,  $G_c(i\omega_n )$  is  given by: 
\bea
 G_c(i\omega_n ) = i \: \frac{sign(\omega_n)}{ \sqrt{ {\cal{J}}} \sqrt{ | \omega _n|}}, 
  \label{galt}
 \enea
From Eq.(\ref{galt1}), with  $\Sigma ( i\omega_n )  \to \Sigma _c^{0}(i\:\omega_n ) =  \sqrt{ {\cal{J}}} \sqrt{ -  i\:  \omega _n}$ in the same limit.
  
  However,  $\partial _\tau$ plays an important role in the extended model of Section III. 
 
 \subsection{quasiparticle energy distribution of the SYK extended model  close to  the Fermi Surface  at  $T \lesssim  T_{coh}$}

  To check the nature of the quasiparticles  for energies close to the  Fermi Surface  (FS), when  the SYK model is extended   perturbatively   to  two space dimensions, we  calculate the quasiparticle occupation number $n_k$  close to the Fermi energy $\epsilon _F \sim  \mu$. 
   
     The single particle  Green's function of the SYK model in the conformal symmetry limit,   The Green function  $G_c(i\omega_n) $  of Eq.(\ref{galt}) is local in space.  To include  the   $k-$dependence   in the propagator, for $k$ vectors close to the FS,
we add a single band energy $\epsilon_{\vec{k}}= \tilde{v}_F k $,  in the continuum limit  to the right hand side of Eq.(\ref{galt1}) and expand to lowest order, obtaining\cite{chowdhurySenthil}: 
\bea
G(\vec{k},i\omega_n ) \approx  \left [\frac{1}{ \Sigma _c^{0}(i\:\omega_n )} + \frac{ \epsilon_{\vec{k}}}{\left |   \Sigma _c^{0}(i\:\omega_n )\right |^2}\right ].
   \enea
   The occupation number  of the complex fermions can be rewritten in terms of the  Majorana fermions  by means of Eq.(\ref{opert}) as: 
     \bea
   n_k = \langle  c^\dagger_k c_k\rangle   = \met  \left [  1 + i\: \langle \chi_2\chi_1 \rangle_k \right ] \hspace{2cm} \label{occup}\\ 
   = \met  \left (  1 +  \frac{1}{\hbar \beta}  \: \sum _n e^{i\:\omega _n 0^+}  \: G( k,i\:  \omega _n)  \right ).\nonumber
   \enea 

   We evaluate  the Matsubara sum  of  $ G( k, i \:\omega_n ) $ in the complex frequency  plane $i \:\omega_n\to z$:
      \bea
 \frac{1}{\hbar \beta}  \: \sum _n e^{i\:\omega _n 0^+} G( k,i\:  \omega _n) \nonumber\\
 =  \frac{1}{ 2 \pi i }  \oint  e^{z 0^+}  \frac{dz}{e^{\beta {\cal{J}} z} +1}\:  G( k, z) ,
     \label{beg}\\
      G( k,z ) \approx  - \frac{1}{\sqrt{{\cal{J}}}} \frac{ sign[z]}{\sqrt{-z}} +\frac{\epsilon _k -\mu}{ {\cal{J}} z }.
      \label{lag}
	\enea
	$ G( k,z ) $ has a cut and a pole at the origin. 
	 The cut can be taken on the real  $z>0$ axis and contributes to the integral of Eq.(\ref{beg}) as follows:
  \bea
  - \frac{1}{ 2 \pi i } \left \{  \int  _{+ \infty}^{0}   \frac{dz}{e^{\beta {\cal{J}} z} +1}\:  \frac{ e^{-i\pi /2}}{\sqrt{z}} 
 +\int  ^{+ \infty}_{0}   \frac{dz}{e^{\beta {\cal{J}} z} +1}\:  \frac{ e^{i\pi /2}}{  \sqrt{z}}\right \}  \nonumber\\ 
= 2  \frac{1}{  \pi  }  \int  ^{+ \infty}_{0} e^{-s0^+}  \frac{ds}{e^{\beta {\cal{J}} s^2} +1} =  - \frac{ ( \sqrt{2} -1 ) \: \zeta \left [\met \right ]}{\sqrt{\pi \beta {\cal{J}}}} > 0,
\nonumber
 \enea
 where  $ \zeta \left [\met \right ] = -1.46035$ is the Riemann function. At zero temperature  the contribution of  the cut vanishes.
 
The second term in Eq.(\ref{lag}) contains  the  pole at the origin. Defining  	 $ \xi = \frac{\epsilon _k -\mu}{ {\cal{J}}}$,  we get  
  \bea
- \frac{1}{ 2 \pi i } \xi \: \lim_ {\eta \to 0^+}  \oint   e^{z 0^+}\:   \frac{dz}{e^{\beta {\cal{J}} z} +1}\:  \frac{1}{ \eta -z}  = - \xi.
 \enea
It follows that  the final result  for Eq.(\ref{occup}), close to the Fermi energy,   in the limit  $\beta {\cal{J}} \to \infty$  and low temperature,  is:  
\bea
n_k = \met \left ( 1-  \frac{\epsilon _k -\mu}{ {\cal{J}}}\right ).
\enea
The FS of the  SYK model  extended  to higher space  dimensionality   is blurred in this perturbative approach and Fermionic quasiparticles are no longer  well defined. 

In Section IV.A  we will  derive the lifetime of the quasiparticles of  a $3-d$ quantum Liquid (qL)  in perturbative interaction with the  $0+1-d$ SYK system  and we  find that the qL  becomes itself a Marginal Fermi Liquid.  
   \subsection{Fluctuations around the IR saddle point solution}
 
 Back to Eq.(\ref{aca}), we can  expand the action to second order and  integrate  out the  $\delta \Sigma $  fluctuations. The resulting functional integral,  in terms of the fluctuations  of the $G$ field, $\delta g(\tau_{12},\tau_+)$, is\cite{noi}
 \begin{widetext}
 \bea
 {\cal{Z}}\left [\partial_\tau\varphi_p(\tau_+) \right ] \propto \int D \!\left ( \Pi \delta g^*_{\tau_{12},\tau_+} \right ) D\!\left (\Pi  \delta g_{\tau_{12},\tau_+} \right )\: 
  e^{\frac{N}{4} \left [\langle \delta g  | K_c ^{-1}-1| \delta g \rangle\right ]} \: e^{-\frac{N}{2} \Re e \left \{ \left \langle - i\: R_c^{-1} \partial _{\tau} \varphi _p | \delta g \right \rangle \right \} }\nonumber\\
  \times e^{\frac{N}{4} \frac{{ t_0}^2}{N}p^2 \left [\langle - i\:\partial _{\tau} \varphi _p  | R_c^{-1}  \Lambda_c  R_c^{-1} | - i\:\partial _{\tau} \varphi _p  \rangle\right ]}
 \label{immf}\\
  g(\tau_1,\tau_2) = R_c(\tau_1,\tau_2)\: G(\tau_1,\tau_2), \:\:\:\:\:  R_c (\tau _1 ,\tau _2)
 = \beta {\cal{J}} \sqrt{3} \left | \tilde{G}_c (\tau _1 ,\tau _2) \right | \nonumber\\
 K_c ( \vartheta _1 ,\vartheta _2, \vartheta _3, \vartheta _4) = R_c (\vartheta _1 ,\vartheta _2) {G}_c( \vartheta _1, \vartheta _3) {G}_c( \vartheta _4, \vartheta _2) R_c(\vartheta _3, \vartheta _4)  \hspace{3cm} \nonumber\\
  = (\beta {\cal{J}} )^2 \: (q-1 ) \left | {G}_c (\vartheta _1 ,\vartheta _2) \right |^{\frac{q-2}{2}} {G}_c( \vartheta _1, \vartheta _3) {G}_c( \vartheta _4, \vartheta _2) \left | {G}_c (\vartheta _3 ,\vartheta _4) \right |^{\frac{q-2}{2}}.\nonumber\\
  \langle  \partial _{\tau} \varphi _p  |  R_c^{-1}\Lambda_c  R_c^{-1}| \partial _{\tau} \varphi _p  \rangle
  = \int d\tau_{12}\int d\tau_+ \left [ \delta G_{c,x} (\tau_{12}, \tau_+) \: \delta G_{c,x'}^*(\tau_{12}, \tau_+)\right ],
  \label{trasf1}\enea
\end{widetext} 
where $q=4$, as  in the usual SYK notation and 
 \bea
  \delta G_{c,x} (\tau_{12}, \tau_+) \delta G_{c,x'}^*(\tau_{34},\tau_+')\hspace{1cm} \nonumber\\
  =
 \left [  G_{x}(\tau_1,\tau_2)\: G_{x'}^*(\tau_3,\tau_4) -G_{c }(\tau_1-\tau_2)\: G_{c }(\tau_3-\tau_4) \right ] \nonumber\\
   \approx \frac{1}{2}\: G_{c }(\tau_{12}) \:\left ( e^{-i   {\tilde a}\cdot  \nabla_x \left [\varphi _x(\tau_+)-\varphi _x(\tau_+')\right ]} -1 \right ) \: G_{c }(\tau_{34})  + c.c. \nonumber
 \enea
 Here we keep only quadratic terms in the phase  difference $\varphi _x(\tau_+)-\varphi _x(\tau_+')$,  because the first order drops out by adding the $c.c.$ and we obtain:
\bea
\to - \frac{1}{2}\: G_{c }(\tau_{12}) \left ( {\tilde a}\cdot  \nabla_x \left [\varphi _x(\tau_+)-\varphi _x(\tau_+')\right ]  \right )\!\!^2\: G_{c }(\tau_{34}).
     \label{pro2}
     \enea
         We now approximate $ \left [\varphi _x(\tau_+)-\varphi _x(\tau_+')\right ]  \approx $ $ \left  (\tau_+-\tau_+' \right )\partial _\tau \varphi _x(\tau_+)$ in the kernel $\Lambda _c$ of Eq.(\ref{trasf1}).  Owing to the
self-averaging established for the SYK model at large $N$,  translational invariance  allows space Fourier transform ($p$ is a wavevector).


     To highlight the physical meaning of the source term  in the functional of Eq.(\ref{immf}), we  define  the energy density current $ \vec{J}^{\cal{E}} $ in a  
      hydrodynamic relativistic  approach\cite{landauHydro}(see Appendix A):
       \bea
 \vec{J}^{\cal{E}} = \frac{( \epsilon +\tilde{p} )}{n} \: \left (\vec J -\vec \nu \right )  = T_0^\alpha \equiv  w \: u^\alpha. 
 \label{cora}
 \enea
Here  $ w=  \epsilon +p $ is the enthalpy with pressure $\tilde{p}$ and  $T_i^\alpha $ is the energy-momentum  tensor (italic indices $ i=0,1,2,3$, while  greek indices $ \alpha =1,2,3 $). 
 $J$ is the average  particle current  density and we will adopt the linear response approach, so that the fluctuating current   $\vec \nu $ is a small contribution, $\vec \nu  << J$.   Here  $u^\alpha$ are the components of the velocity and  $u^0 =1$. 
 
  The starting point  of Appendix A is the conservation of the 
energy-momentum tensor $T_i ^k$, which  includes dissipative processes.  The  viscosity and  the thermal dissipation are  accounted for by   the stress tensor  of Navier-Stokes origin,  $\tau _i^k$. Here   $ u^i \tau _i^k =0 $, because  the dissipative components of the stress-energy tensor and current are orthogonal to $\vec{u}$. In the absence of external forces and in  the case of $J=0$,  it can be  shown in Eq.(\ref{cruk}) that the change of entropy (per unit volume) is given by:
  \bea
     T\:   \frac{D s}{Dt} = \mu  \nabla \nu  - \tau _i ^k \frac{\partial u^i}{\partial x^k} \nonumber\\
     = - \nu  \cdot \nabla  \mu  + \nabla  \cdot \left ( \mu  \vec{\nu} \right )  - \tau _i ^k \frac{\partial u^i}{\partial x^k}.
\label{cruk1}
\enea 
 Here $\frac{D}{Dt} =  u^i   \frac{\partial }{\partial x^i}$, is the covariant  derivative, which  includes the convective derivative.
The last two terms on  the right hand side are dissipative terms.
 When we integrate over space and time  and we  substitute  $ \nu /n  \to  \vec{J}\: ^{\cal{E}} /k_BT $  from Eq.(\ref{cora}),  we get a source term to be added to the action of Eq.(\ref{tron}).
     \bea
     [\delta S]_{source} = \int dt\int dv_b  \:\:\frac{1}{k_BT}  \vec{J}\: ^{\cal{E}}  \cdot \vec{\nabla} \mu,
     \label{essource}
     \enea 
     which reminds of the source term  $\propto \int J\cdot E$ of electrodynamics. 
     $T$ plays here the role of the background average  constant temperature in linear response and    $v_b$ is a small volume.  As $ \mu(x,\tau_+) = \hbar\: \partial _{\tau_+} \varphi _x$,  it follows that  fluid   correlations  contributing to energy density currents,   $\frac{1}{(k_BT)^2 } \langle \vec{J} ^{\cal{E}}_p\:\vec{J}^{\cal{E}}_p\rangle \:  \equiv   \langle \dot{\cal{N}}_p(\tau_+) \dot{\cal{N}}_p(0)  \rangle $, are generated by the functional of Eq.(\ref{tron}). This allows us  to discuss  thermal transport in the extended SYK model within the conformal limit, what is done in the next Subsection.
     
          Meanwhile, it is found that the approximate conformal symmetry of the SYK model is spontaneously broken down to the $\wt {SL}(2, \RR)$ group symmetry\cite{kitaevGeom}. The dominant contribution to the functional integration in Eq.(\ref{immf}) comes from the pseudo Goldstone modes (pGm)  $\delta g_{\tau_{12}}$. At this  level of approximation, the  $\delta g$'s  do not depend on $\tau_+$   and are eigenfunctions of the Kernel $ K_c$ with eigenvalue 1.  The signature of the symmetry breaking  is  the disappearance of 
 the action of the $\delta g$'s $\propto \langle \delta g  | K_c ^{-1}-1| \delta g \rangle $  in  Eq.(\ref{immf}), what implies  that  the functional integral is actually divergent. Regularization requires  ultraviolet (UV) corrections.   In the next Subsection we are going first to discuss transport in the conformal symmetric limit by considering the saddle point for the action only, which suffices for a linear response approximation. The divergence of the functional integration will be cured in Section  III. 
     
     \subsection{Transport in the  IR  conformal symmetry limit}
 In the conformal symmetry limit,  we  keep only the  lowest order in the kinetic term by  approximating   $ G_{x}(\vartheta_1,\vartheta_2) \approx  {G}_c( \vartheta _1, \vartheta _2)\: e^{i \varphi _x (\tau_+)} $ in Eq.(\ref{greo}). The kinetic term operator appearing in  Eq.(\ref{trasf1}), $ \frac{1}{2}\frac{ p^2 {\tilde a}^2}{\hbar^2}  R_c^{-1}\Lambda_c  R_c^{-1}$,  is the  space Fourier Transform  of
 \bea
  \frac{1}{2} ( {\tilde a}\cdot   \overset{\leftarrow}{ \nabla}_x  )G_{c }(\tau_{12})(\tau_+-\tau_+')^2 G_{c }(\tau_{34}) ( {\tilde a}\cdot   \overset{\rightarrow}{ \nabla}_{x'} )
  \label{dit}
  \enea
     derived from Eq.(\ref{pro2}).
    The function  $\hbar \partial _{\tau} \varphi _p $ is canonically conjugated to the particle flux  $\dot{{\cal{N}}}_p(\tau_+) =\frac{1}{\tau_0} \frac{\delta I_{ex} }{\delta \hbar \partial _{\tau} \varphi _p }$,  where  $ I_{ex} [\partial _{\tau} \varphi _p ]$ originates from   
    the action of Eq.(\ref{aca})  and   we have defined  $ \tau _0 = \sqrt{ \langle \left ( \tau _+- \tau _+'\right )^2\rangle} $ as an average timescale of the correlations in  time, according to Eq.(\ref{dit}).  The  function $ \frac{ \delta}{\delta \dot{ \varphi} } \langle  \dot{\cal{N}}_p(t) \rangle$ and the current-current   correlator $\langle \dot{\cal{N}}_p(t) \dot{\cal{N}}_p(0)  \rangle$ can be derived from  the second variation  of the generating functional of Eq.(\ref{immf}),  $ \frac{\delta^2}{\delta 	{\dot {\varphi}} ^2} \ln   {\cal{Z}} $.   Moving back to space  and  real time, the   correlator in the hopping between sites, at saddle point,  is
    \bea
    \frac{t_0^2}{2 \hbar }  \langle \dot{\cal{N}}_p(t) \dot{\cal{N}}_p(0)  \rangle  \to  \frac{t_0^2}{2 \hbar } \left ( i\: G_{cL}^\beta (t) \right ) \left (- i\: G_{cR}^\beta (-t) \right ),
    \label{corno}
    \enea
    where  the labels $L,R$ refer to left and right side in the hopping process and we have lumped in $t_0$ all prefactors. 
    Here
  \bea
i \: G^\beta (t) = \left  \{ \begin{array}{cc} \int d\omega  \: A_+ (\omega) \: e^{-i\: \omega t}  &  {\rm for} \: \:  t >0  \\
 -\int d\omega  \: A_- (\omega) \: e^{-i\: \omega t}  &  {\rm for} \: \:  t <0 \end{array} \right .\nonumber\\
\enea
where  $ A_\pm$ are the fermionic spectral functions with   $  A_+^\beta ( \omega) =  e^{\beta \omega}\:\:   A_- ^\beta( \omega) $.  The $A_\pm$'s are real and positive. 

 Let $
\langle  \dot{\cal{N}}(t) \rangle _{\substack{\\ \leftarrow\\\rightarrow}}$ be  the currents to the Left/Right,  which  differ for a
 chemical potential $\mu$  imbalance. From Eq.(\ref{corno}), we have: 
\bea
 \frac{ \delta}{\delta \dot{ \varphi} } \langle  \dot{\cal{N}}(t) \rangle _{\substack{\\ \leftarrow\\\rightarrow}} =    -\frac{1}{4 \hbar}t_c^2  N\: \int d\omega \: \int d\nu \: \left [ A_{+L}(\nu )\: A_{-R} (\nu \mp \omega) \right . \nonumber\\
\left . -A_{-L} (\nu  ) \: A_{+R}(\nu \mp\omega )\right ] \:  e^{\pm i\: \omega t}, \:\:\:   {\rm for} \: \:  t >0 . \hspace{1cm}\label{suscb}
 \enea
 The spectral functions are actually  independent of the side: $ A_{\pm L}^\beta(\omega ) = A_{\pm R}^\beta(\omega )$.
  
    According to Eq.(\ref{essource}), the  difference between Left and Right fluxes over $\omega $,  in the limit $\omega \to 0$, is connected to  the response function to an external perturbation of the chemical equilibrium:
     \bea 
      D^{R\beta}(\omega) =  -i\: \int dt\:  \theta (t) \left   \langle \left [ \dot{\cal{N}}(t), \dot{\cal{N}}(0) \right ] \right \rangle  \: e^{ i\: \omega t},\nonumber
      \enea
      with
      \bea
    \left   \langle \left [  \dot{\cal{N}}(t), \dot{\cal{N}}(0) \right ] \right \rangle  \approx  \frac{t_0^2 }{4 \hbar } \left [ \left ( i\: G_L^\beta (t) \right ) \left (- i\: G_R^\beta (-t) \right )\right .
  \nonumber\\
\left . - \left (- i\: G_L^\beta (-t) \right ) ^* \left ( i\: G_R^\beta (t) \right ) ^*\right ]. \label{susc}
 \enea
Indeed, by comparing with Eq.(\ref{suscb}), 
 \bea
 \lim _{\omega \to 0}\left [ \frac{ \delta  }{\delta  \dot{ \varphi} } \: \left \langle \left (\dot{\cal{N}}_{\leftarrow} -\dot{\cal{N}}_{\rightarrow}\right )\right \rangle\right ]_ \omega\nonumber\\
  = -   \frac{  t_0^2 \tau_0 }{2\pi^2\hbar} N \:  \int d\nu \:\Im m \left \{ D^{R\beta}(\nu)\right \} \: \frac{ \partial }{\partial \nu }   \Im m \left \{ G^{R\beta}(\nu)\right \} \:\:\:
  \label{prep}
 \enea
 provides 
 the real part of the conductivity, with the density of  the incoming states $ - \Im m \left \{ G^{R\beta}(\nu)\right \}/\pi  $.

         The thermal conduction arises from    the response to the energy current in place of the particle current.  It is enough to substitute $ A(\nu  ) \to  \hbar \nu \: A(\nu  ) $ in the final expression to obtain  the thermal conduction response $Re\{ \kappa \} $. On the other hand, at equilibrium, 
 the imaginary part of the response function $  D^{R\beta}(\omega)$  and the retarded function $  G^{R\beta}(\omega)$ are simply related by
  \bea
  \Im m \left \{ D^{R\beta}(\omega)\right \}  = \tanh \frac{\beta \omega}{2} \:   \Im m \left \{ G^{R\beta}(\omega)\right \}.
  \label{fdis} 
  \enea
  We obtain
       \bea
\frac{\kappa}{k_B} =   \lim _{t \to \infty}\left [\frac{Re\{ \kappa  (t)\} }{k_B}\right ]  =   \lim _{\omega \to 0}  |\omega| \: \Re e \left \{ \kappa ( \omega ) \right \} \hspace{1.5cm}\nonumber\\
 =N\:  \frac{t_0 ^2 }{2 \hbar }  \int d\nu  \: {\rm tanh}\frac{\beta \nu }{2} \: \frac{\partial}{\partial \nu } \left [ -\frac{\nu   }{\pi}  \:  \Im m \left \{ G^{R\beta}(\nu )\right \} \right ]^2.\nonumber
 \enea
 The SYK Green function in the conformal limit which was approximated in Eq.(\ref{galt}), is:  
 \bea 
 G^{R\beta } (   \omega  ) = - i\: b\:\beta\:    \frac{1}{\sqrt{\pi} }\: \frac{ \Gamma \left (  \frac{1}{4} - i\: \frac{\beta \omega  }{2 \pi } \right )}{ \Gamma \left (  \frac{3}{4} - i\: \frac{\beta \omega  }{2 \pi } \right )}, \:\: b = \frac{ \pi ^{1/4} }{\sqrt{2\beta {\cal{J}}} } 
  \enea 
and, integrating by parts, we get: 
 \bea
  \lim _{t \to \infty}\left [\frac{Re\{ \kappa \} }{k_B}\right ]_{t}  =  \frac{1}{\pi^{3/2} \hbar }  \:  t_0^2 \frac{N}{2 {\cal{J} } }  \hspace{2.5cm}  \label{kppa}
\\
\times   \int dx \: {\rm sech}^2  ( \pi x)  \:\left [  \Re e \left \{ \frac{ \Gamma \left (  \frac{1}{4} - i\: x\right )}{ \Gamma \left (  \frac{3}{4} - i\: x \right )}\right \}\right ]^2 \propto  N \:\frac{ k_BT_{coh}}{\hbar },
\nonumber
\enea
where we have introduced the definition of $T_{coh}$.  

 By the same token, from Eq.s(\ref{susc},\ref{prep}), the particle current can be derived following  the same steps as before. Ignoring the fact the current is neutral in this Majorana fermion system, an hypothetical  electrical real conductivity  $\sigma $ would be\cite{song}: 
 \bea
  \lim _{\omega \to 0}\left [\frac{Re\{ \sigma \} }{e^2}\right ]_{\omega}  = N\frac{\beta}{\pi^{3/2}}  \: \frac{ t_0^2}{{\cal{J}} }\hspace{2.5cm}\nonumber\\
 \times  \int dx \: {\rm sech}^2  (\pi x)  \:\left [  \Re e \left \{ \frac{ \Gamma \left (  \frac{1}{4} - i\: x\right )}{ \Gamma \left (  \frac{3}{4} - i\: x \right )}\right \}\right ]^2 \propto N  \frac{ T_{coh}}{T}.
 \label{condur}
\enea
Eq.(\ref{condur}) is   the celebrated linear $T $ dependence of the   electrical resistivity offered by the SYK model.   In the next Section we discuss the contribution of the collective  Q-modes to the  $T-$dependence of the transport coefficients. These  modes  arise when  the UV corrections to the action are included.

 \section{ collective  bosonic excitations from UV  corrections}
 In the previous Section, based on  the spontaneous conformal symmetry  breaking,  we acted as  including just the Goldstone modes in the functional integration  and regarded    Eq.(\ref{immf}) as   a generating functional for  the Goldstone mode correlations, having   $\partial _{\tau_+} \varphi _p  $  as  a source field: 
   \bea
     {\cal{Z}}\left [ \partial _{\tau} \varphi _p \right ]  =\int D   \delta g_{\tau_{12}} \:   \: e^{-\frac{N}{2} \left \langle - i\: R_c^{-1} \partial _{\tau} \varphi _p | \delta g \right \rangle}
\nonumber\\
 \times  e^{\frac{N}{4} \frac{{ t_0}^2}{N}p^2 \left [\langle - i\:\partial _{\tau} \varphi _p  | R_c^{-1}  \Lambda_c  R_c^{-1} | - i\:\partial _{\tau_+} \varphi _p  \rangle\right ]}.
 \label{tron}
    \enea
We have swept under the carpet the fact that  these  massless Goldstone modes  are responsible for the divergence of the functional integral of Eq.(\ref{immf}).  Moreover,  $\varphi _x(\tau_+)$  was given as an external source with no dynamics. However,  symmetry breaking  is only approximately spontaneous, because  the term  $\sim \partial_\tau$  in Eq.(\ref{aca})  produces itself   a  symmetry breaking,  which becomes relevant in the UV.   The modes of the gaussian action are, so to say, pseudo Goldstone modes (pGm) and display a small mass.  In the IR,  a  time  reparametrization  of the conformal Green's function under the diffeomorphism   $ e^{i\vartheta }\to e^{i \:  \varphi _x(\vartheta )}$ produces a  local term to the action, which gives dynamics to the  $\varphi _x(\tau_+)$  fluctuations and generates  diffusive  collective bosonic excitations which we call Q-excitations. 
  As the $0+1-$SYK dots at the lattice sites are sinks of charge and momentum, the surviving dynamical variable is energy  density and we interpret  the  Q-excitations as   energy density currents.  
   
   The IR Green's function  under  reparametrization reads ($\varphi '\equiv \partial_\vartheta \varphi$, $\Delta = \frac{1}{4}$):
    \bea
  \tilde{G}_{IR}  (\vartheta _1 ,\vartheta _2)  = \tilde {G}_c \left (\varphi(\vartheta _1),\varphi(\vartheta _2) \right )\: \varphi '(\vartheta _1)^{\Delta },\varphi '(\vartheta _2)^{\Delta }, 
  \nonumber
  \enea
  which, when expanded to lowest order in  $\vartheta_{12}$, gives 
  \bea
    \approx   \tilde{G}_{\beta \to \infty} \!\! (\vartheta _1 ,\vartheta _2)\! \left [ \!1+ \!\frac{\Delta }{6} (\vartheta _{12})^2 Sch \!\left (e^{i\: \varphi(\vartheta _+)},\!\vartheta _+\right )\right ],
       \label{schvario}
     \enea
     where $ Sch \left (  e^{i\: \varphi(\vartheta _+)}, \vartheta _+\right )$ is the Schwarzian\cite{kitaev}.  The fluctuations $ \delta g_{\tau_{12}}  \sim  R_c \left (  \tilde{G}_{IR} -  \tilde{G}_{\beta \to \infty}\right ) $  can be substituted in the source term of Eq.(\ref{tron}), giving rise to a local   action of order $1/ \beta {\cal{J}} $:
   \bea
    \frac{ I_{local}}{N} [\partial_\tau \varphi]= -2\pi \:\frac{ \alpha _S }{\beta {\cal{J}}} \!\int \frac{d\vartheta}{2\pi} \left [ \met - \frac{ ( \varphi '')^2 - \!( \varphi ' ) ^2}{2} \right], 
    \label{schwarz}
       \enea
      which  adds   dynamics  to the field  $\varphi _x(\tau_+)$. The constant term in Eq.(\ref{schwarz})  gives a contribution  $\propto \beta $, which  is just a correction to the energy and will appear in Eq.(\ref{correneg}).    The added local action of Eq.(\ref{schwarz})  is $ {\cal{O}}\left (( \beta {\cal{J}}^{-1})\right ) $ and a correction  of the same order should be included to the action\cite{maldacena,noi}, coming from the $\langle \delta g  | K_c ^{-1}-1| \delta g \rangle$ term appearing  in Eq.(\ref{immf}).  We remind that $\partial_\tau \varphi _x(\tau_+)$ corresponds to  changes in the chemical potential $\mu$, according to Eq.(\ref{essource}). 
      
The  UV correction  of the $0+1-d$ SYK model is local in time.  To extend  the  model  to higher space dimensions   we use  the macroscopic continuity equation
  \bea
    \rho \: T  \: \frac{D s }{Dt}  = - \nabla \cdot J^{\cal{E}}, 
    \label{eq01}
    \enea
    where  $ J^{\cal{E}}$ is the energy current density of Eq.(\ref{cora}). 
  This will help us in characterizing  the collective bosonic Q-excitations.
     In the next Subsection we derive the response function to $ J^{\cal{E}}- J^{\cal{E}}$  fluctuations, starting from the equation of motion, which is assumed to be diffusive:
        \bea
   J^{\cal{E}} = -\kappa \: \nabla T + g, 
   \label{diro}
   \enea
   where $g$  is an  external stochastic source. 
    Diffusivity, besides being physically justified in the incoherent regime, can also  be justified starting from the full action $I[ \partial _\tau \varphi_p] $ in the large ${\cal{J}}$ limit\cite{noi}. 
  
 \subsection{Hydrodynamic response to energy current density fluctuations }
 According to Eq.(\ref{essource}), the gradients  of the chemical potential  can be obtained as a response to energy current density fluctuations in the $2+1-d $ SYK  system. This is of importance for us, as the  perturbation of the local chemical potential  in the  $3-d$ qL,  induced by the energy current density fluctuations of the $2+1 -d$ SYK  system,  is our final goal.  While in Section II.C we have presented the response of  the energy current  density $J^{\cal{E}}$ to chemical potential gradients $\nabla \mu$,  in this Subsection we study the response of the chemical potential to the $ J^{\cal{E}}$  fluctuations,  starting from temperature gradients and from the continuity equation, Eq.(\ref{eq01}).
 
 Let us define the  canonical conjugate variables  $ x_a, X_a$,  appearing in the  continuity equation, Eq.(\ref{eq01}):
     \bea
 \dot{S} = - \dot{x}_a \frac{\partial S}{\partial x_a}, \:\:\:  {\rm with } \:\:   X_a = - \frac{\partial S}{\partial x_a},
 \label{conj}
\enea
in terms of which  the  usual linearized response equation takes the form:
  \bea
 \dot x_a = \sum _b \left ( \gamma_ {ab} X_b  + g_b \right ) 
      \label{eq12} 
  \enea
(here  $ g_b$ is the  external stochastic source and $a,b$ denote  elementary  space cells of volume ${v}_a, {v}_b $).
 
Space integration of Eq.(\ref{eq01}), together  with Eq.(\ref{diro}), gives:
\bea
  \frac{\dot S}{k_B}  = \int  d v_b \: \frac{1}{k_B T} \: \nabla \cdot \left ( \kappa \nabla T \right ) \hspace{2cm}\nonumber\\
   = - \int d{v}_b \: {\left \{ \frac{ J^{\cal{E}}}{k_B}\right \}_b \: \left \{\frac{ \nabla  T }{T^2}\right \}}_b.
  \label{doti}
  \enea
  It follows from Eq.s(\ref{conj},\ref{eq12},\ref{doti})  that  
   \bea
   \dot x  \sim g  =   \frac{ J^{\cal{E}}}{k_B }, \:\:\:  X= \frac{ \nabla  T }{T^2}, \:\:\:   \gamma _{ab}= - \frac{ \kappa\: T^2}{k_B }.
\label{mot}
 \enea
The hydrodynamical approach developed in  Appendix A, provides Eq.(\ref{cruk}), which reads:  
 \bea
  \frac{D s}{Dt} - \frac{\mu }{T}  \nabla \nu  =  T \: \nu \!\cdot \! \nabla \frac{\mu }{T}  - \tau _i ^k \frac{\partial u^i}{\partial x^k},
\label{cruki0}
\enea
with $\nu$ defined in Eq.(\ref{cora}). 
When the particle current vanishes, the right hand side contributes with  dissipative terms which can be cumulatively expressed as $ -\frac{\partial w}{\partial t}$. The term $ \frac{\mu }{T}  \nabla \nu $ provides 
 the diffusive contribution to the change in time of the entropy. Eq.(\ref{moti}) shows that $ - \frac{w}{n} \approx - \kappa \nabla^2 T$. Introducing the diffusive constant  $ \tilde {D} = \kappa / nC$, where $C$ is the specific heat, and posing $ C \nabla^2 T \approx \nabla^2 w$,  we obtain:
 \bea
 k_B \: \dot x_b\: X_b  =  - \frac{D s_b}{Dt} =   \frac{n}{w}  \frac{\mu }{T} 
  \left ( \tilde{\cal{D}}
  \:  \nabla ^2 w -\frac{\partial  w}{\partial t} \right )_b. 
  \label{nonco}
  \enea
  Fourier transforming in space and inverting this equation, we get the response equation that we are looking after ($ \mu  \to \mu(\omega \approx 0 )$):
  \bea
 \mu_{q, \omega } \: n= { k_BT }  \: {x}_{b,\omega } \: 
  \frac{-i\omega }{ \tilde{\cal{D}} \: q^2 -i \omega}\: X_{b,-\omega}. \hspace{6cm} 
  \nonumber
  \enea 
  The source $X_{b,-\omega}$ in this equation can be interpreted  as an elementary   energy current density  source:  $  {j^{\cal{E}} _{-q ,-\omega}}  = \frac{T k_B^2}{h}  [ \nabla _{x'} T  ]_{q }$. In fact $\kappa _0=  \frac{T k_B^2}{h} $ has the dimension of a thermal conductance\cite{pendry} $ [C/time]$. We obtain: 
   \bea
 \mu_{q, \omega } \: n=  \frac{h  }{(k_BT)^2} \:   j^{\cal{E}} _{-q ,-\omega} J^{\cal{E}}_{q ,\omega}\:   \frac{-i\: \omega  }{ \tilde{\cal{D}} \: q^2 -i \omega}. 
   \label{muo}
  \enea 
  In the following we  consider  the thermal average of $ \int d\omega ' J^{\cal{E}} _{-q ,\omega -\omega'} J^{\cal{E}}_{q ,\omega'}\:  $, as inducing a response on the space and time dependent chemical potential energy density $ \mu \: n$. 
 We will use this result when  discussing  the interaction between the low energy qL and the bosonic Q-excitations.  In the Subsection B,  the thermal average $ \int d\omega '  \left \langle \left \{J^{\cal{E}} _{-q ,\omega -\omega'} J^{\cal{E}}_{q ,\omega'}\right \} \right \rangle$ will be denoted as  $ \left   \langle  \left \{ J^{{\cal{E}}}_{-q} ,J^{{\cal{E}}}_{q} \right \}  \right  \rangle_\omega$.

Classically, the symmetrizied correlation derived from  from Eq.(\ref{eq12}) is\cite{weiss}  $\met  \left \langle  \left \{ g_a(t) ,\: g_b(t' ) \right \} \right \rangle \:v_b =\left ( \gamma _{ab} + \gamma_ {ba} \right ) \: \delta ( t-t') $  and, with the correspondence of Eq.(\ref{mot}),   it becomes  in     the quantum case:
    \bea
  \met \int d^2r \int dt\int dt'\: \left  \langle  \left \{ J^{{\cal{E}}}(r,t) ,J^{{\cal{E}}}(r,t')  \right \}  \right \rangle  \: e^{i\:\omega  (t-t')} \nonumber\\
=- \hbar \omega  \:T \kappa \:  \coth\left (\frac{ \hbar \omega}{2k_B T }\right ). \hspace{2cm}
  \label{vir13} 
   \enea
   where $\kappa \sim \lim _{t\to \infty} Re\{ \kappa \} _t= \lim _{\omega  \to 0}   |\omega| \: \Re e \left \{ \kappa ( \omega ) \right \} \: $, so that, substituting the  $J^{{\cal{E}}}-J^{{\cal{E}}} $ fluctuations with the imaginary part of the response via the fluctuation dissipation theorem, we have: 
   \bea
   \omega \:  \Re e \{\kappa (\omega)\} =   \met\: \Im m \left \{D^\beta _{ J^{{\cal{E}}},J^{{\cal{E}}}}\!(\omega ) \right \}\! \frac{1}{\hbar \omega  T}\! \tanh\left (\frac{ \hbar \omega}{2k_B T }\right ), \nonumber\\ 
   \label{vir14}
   \enea
 where 
  \bea
   D^\beta _{ J^{{\cal{E}}},J^{{\cal{E}}}}  (\omega )     = -i     \int dt\:  \int dt'\:  \Theta(t-t')\: \nonumber\\
\times   \int d^2k \left  \langle \left [J^{{\cal{E}}}_{-k}(t) ,\: J^{{\cal{E}}}_{k}(t' ) \right ]\right \rangle \: e^{i\:\omega  (t-t')}. 
\label{res}
   \enea
   The response function of Eq.(\ref{res})  will be evaluated in the next Subsection  for damped  Q-excitations within a restricted  $2-d$ volume in which correlation is non vanishing. 
   The  space volume integration is explicit in Eq.(\ref{vir13}) for the  $J^{{\cal{E}}}-J^{{\cal{E}}} $ fluctuations  and implicit in the $k-$integral of Eq.(\ref{res}). We will denote this restricted  $2-d$ volume  as $ \tilde{a}^2_\ell $ in Subsection B. 
       \subsection{Lagrangian  approach to gapless diffusive  excitation modes}
   
Eq.(\ref{diro})  is a classical diffusion equation of a non conserving system. We now construct a Lagrangian  of  excitation modes  which is conserving but, assuming a relaxation time $\tau _0$ for these modes, it reproduces diffusive motion. We will quantize this Hamiltonian  and derive the fluctuations of these modes from  the response function by  means of  the fluctuation-dissipation theorem.   The canonical conjugate variables and the corresponding   Lagrangian (in 2-d) are:
 \bea
 \dot \theta = \left ( \frac{\kappa C}{\hbar T}\right )^{1/2} \: \frac{ J_{\cal{E}}\tau_0}{k_B},\:\: \label{can}\\
  \nabla  \theta = \left ( \frac{\hbar}{\kappa C\: T}\right )^{1/2} \: \frac{\kappa}{T}   \nabla T\nonumber\\ 
 {\cal{L}}=  \met \int d^2 x\: \left [\frac{k_B}{T}\: \left (\frac{ J_{\cal{E}}\tau_0}{k_B}\right )^2 
 +   \frac{\hbar }{\kappa C} \: \left (\frac{ \kappa}{T}  \nabla T \right ) ^2\right ]\nonumber\\
  \equiv  \met  \int d^2 x\: \left [ A\:  \dot \theta ^2+ B \:  (   \nabla  \theta )^2  \right ] \label{lagx}\\
  A = \frac{\hbar k_B}{\kappa C}, \:\:\:  B=T. \nonumber
 \enea
  Note that the dimension of the two terms in square brackets is $ [ {\cal{E}}/\ell ^2 ]$.

 From the equation of motion:
 \bea
 \frac{d}{dt} \left (\frac{\partial L}{\partial \dot \theta }\right ) -\frac{\partial L}{\partial \nabla \theta }
 = A \ddot \theta + B \nabla \theta =0,
 \enea
 the classical motion equation  of Eq.(\ref{diro}) can be recovered if we put  $ \tau _0 \dot {J}_{\cal{E}} \approx   {J}_{\cal{E}}$. 

  Although $\hbar$ is already in the Lagrangian, we proceed to quantize  the  theory.  Fourier transforming the Lagrangian:
 \bea
  {\cal{L}}=  \met \sum_{k} \: \left [ A\: { \dot \theta}_{-k} { \dot \theta}_k - B \: k^2   \theta _{-k} \theta _k  \right ],\hspace{1cm} 
  \label{lagk}
 \enea
 the canonical momentum for $\theta _k$ is $ \pi_k = \frac{ \partial  {\cal{L}}}{\partial \dot \theta _k} = A \dot \theta _{-k}$, and  $\left [ \theta _k, \pi _{k'} \right ] = i\: \delta _{ k k'}$. It follows that
  \bea 
  \pi _k = -i\: T^{1/2} \frac{1}{  (2 \eta_k)^{1/2}} |k| \: \left ( \hat a_{-k} - {\hat a}^\dagger _k \right ),\nonumber\\
 \theta _k =  T^{-1/2} \frac{  (2 \eta_k)^{1/2}}{ |k|}  \: \left ( \hat a_{k} + {\hat a}^\dagger _{-k} \right ),\nonumber\\
  {\cal{H}} = \sum _k   \eta_k \:  {\hat a}^\dagger _{k}\hat a_{k}  + cnst , \:\:\:  \eta _k = \tilde{a} \left [\frac{ \kappa }{k_B}  \frac{ C T}{\hbar }\right ]^{1/2}\!\!\!\!  |k|  \equiv v\:  |k| , \nonumber\\
  \label{coi12}
  \enea
  what defines the velocity $v$ in terms of a lattice parameter $\tilde a$.  
  The approach is similar to the one for phonons, but  $\pi (x) $ plays the role of the lattice displacement  $d(x,t )$, while $\nabla \theta $ plays the role of  the corresponding  momentum $\pi _{ph}(x,t) $. The atomic mass  $M$ of the phonon problem corresponds to $1/T$ here. With the definitions given  in  Eq.s(\ref{can},\ref{coi12}) 
 \bea
 \met   \left   \langle  \left \{ J^{{\cal{E}}}_{-k} ,J^{{\cal{E}}}_{k} \right \}  \right  \rangle_\omega = \frac{1}{\tau _0^2}  \frac{ \kappa CT}{2 \hbar} 
  \left   \langle \left \{ \pi_{-k} ,\pi _{k} \right \} \right \rangle_\omega. 
    \label{simma}
    \enea
  The space integral of the   response function, $ D^{\beta}_{\pi\pi}(\omega ) $, corresponding to   Eq.(\ref{res}), but  referring to the symmetrized correlation $ \left   \langle \left \{ \pi_{-k} ,\pi _{k} \right \} \right \rangle_\omega $,  can be expressed in terms of the convolution in $k$  space.  In turn, the latter can be transformed into an integral over the mode energy $\eta _k$ defined in Eq.(\ref{coi12}). To account for thermalization, we  introduce   the inverse  lifetime of the collective mode appearing  in Eq.(\ref{can}), as a damping of the mode, $ \Gamma= \hbar /\tau _0 $ in the response function:  
            \bea
    D^{\beta}_{\pi\pi}(\omega )  = \frac{ \pi  T}{v^4\hbar}  \hspace{3cm} \nonumber\\
\times\:  \tilde{a}_\ell^2  \int_0^{\infty} \eta ^2 d\eta\: \left ( \frac{1}{\omega + \eta +i\: \Gamma}  -  \frac{1}{\omega - \eta  +i\: \Gamma} \right )\: e^{-\eta 0^+}
\label{rito}
\enea
so that, 
\bea
\Im m \left \{  D^{\beta}_{\pi\pi}(\omega )\right \}\to  - \frac{2 \pi  T }{v^4} \: \tilde{a}_\ell^2  \: \omega  \frac{\Gamma}{\hbar}  {\rm \:\:\:\:\:\:  for \:\:  \Gamma /\omega  <<1 } . \label{imo}
\enea
Here  we have assumed that the correlations  range a space  area $\tilde{a}_\ell^2 >>\tilde{a} $. 
Using Eq.(\ref{simma}), we get   
 \bea
      \met \:  \tilde{a}_\ell^2  \int k dk   \left   \langle \left \{ J^{{\cal{E}}}_{-k} ,J^{{\cal{E}}}_{k} \right \}  \right  \rangle_\omega = - \hbar \:  \omega \: T    \: \coth\left (\frac{ \hbar \omega}{2k_B T }\right )
      \nonumber\\
      \:\times  \left [  \frac{ \tilde{a}_\ell^2 }{\tau _0^2}  \met  k_B \: 
     \frac{2 \pi   }{v^2} \frac{\Gamma }{\hbar} \right ]. \label{cocori}
     \enea
Comparing   Eq.(\ref{vir13}) with Eq.(\ref{cocori})  we see that the square bracket provides an expression for  $\kappa$: 
\bea
  \kappa  =  \frac{ k_B \Gamma }{\hbar } 
 \left (  \frac{ \tilde{a}^2_\ell }{ \tilde{a}^2}\: \frac{\pi\Gamma  } { C T } \right )^{1/2}, 
 \label{conde}
\enea
where the velocity given by Eq.(\ref{coi12})  has been inserted.  

 The diffusivity $ \tilde{D}_Q $   defined in Eq.(\ref{ded})  is connected to the thermal conductivity $\kappa$  by the Einstein relation: 
  \bea
  \tilde{D}_Q  = \frac{ \kappa}{\rho_0 C} . 
  \label{caco}
  \enea
The   "particle density" $\rho_0$  appearing here is  associated to  the Q-excitations and 
   can be  defined consistently, on the basis of the parameters used in the hydrodynamical model of this Section, as   $\rho _0= 1/v^2\tau_0^2$. Inserting  Eq.(\ref{conde}) in Eq.(\ref{caco}) and using   $\rho_0$  with the velocity given in Eq.(\ref{coi12}),   gives:
   \bea
     \tilde{D}_Q  =  \tilde{a}^2 \: \pi^2 \Gamma \:   \frac{ k_B T}{\hbar } \: 
 \frac{ \tilde{a}^2_\ell }{ \tilde{a}^2}\: \frac{1 } { C T }. 
 \label{ronde}
\enea
 As the energy diffusion is mainly due to heat transport in a highly thermalizable  environment, an estimate of  $CT \: \tilde{a}^2/\tilde{a}^2_\ell $,  appearing in Eq.s(\ref{conde},\ref{ronde}), can be derived from the saddle point contribution to  the energy of the SYK model\cite{maldacena}: 
  \bea
      \ln {\cal{Z}} = -\beta E_0 + S_0 + \frac{c}{2\beta} + ... ,\:\:\:  \frac{c}{2}=  \frac{ 2\pi^2 \alpha _S N}{{\cal{J}}}.
       \label{correneg}  
       \enea 
       By posing  $ \frac{ \tilde{a}^2}{ \tilde{a}_\ell^2 } CT\sim  \partial _\beta \ln Z = {c}/{2\beta^2}$, we get 
   \bea
 \tilde{D}_Q 
  =\frac{ \Gamma }{\hbar} \left ( \frac{ \beta {\cal{J}}} { \pi \alpha _S N }  \right ) \: \tilde{a}^2 .
  \label{dif1}
  \enea
 Eq.(\ref{dif1}) allows to define  $ \tilde{a}_\ell^2 $ as the  mean squared diffusive length, such that  $ \tilde{D}_Q \tau_0 \sim \tilde{a}_\ell^2 $. In turn,  a temperature $T_0$ can be defined, 
  \bea
 \tilde{a}_\ell^2 =   \tilde{a}^2\left ( \frac{ \beta {\cal{J}}} { \pi \alpha _S  N}  \right ), \:\:  T_{0} \sim \frac{\Gamma}{k_B} \left ( \frac{ \beta {\cal{J}}} { \pi \alpha _S  N}  \right ).
 \label{tdisst}
  \enea
   $T_{0} $  is   the  threshold temperature for   thermalization to be efficient  and is independent of temperature if $\tau _0 \sim \hbar \Gamma^{-1}   \sim   \hbar \beta $. Indeed, a "Planck"  relaxation time scale  is expected when the  full incoherent  SYK regime   sets in\cite{hartnollNatPhys,noi}.   In this case,   both   $\tilde{D}_Q $  of  Eq.(\ref{dif1}) and  $T_{0}$ of   Eq.(\ref{tdisst}) are  temperature independent.

   Eq(\ref{tdisst})  confirms that $T_{0}$ is  ${\cal{O}}\left ( \frac{ \beta {\cal{J}}}{ N }\right )$, if just the zero order for $\Gamma$ is retained.   If we make the same  substitution for   $CT \: \tilde{a}^2/\tilde{a}^2_\ell $ in Eq.(\ref{conde}) that  we did in Eq.(\ref{dif1}),  we get: 
    \bea
 \kappa  =  k_B \: \frac{\Gamma}{ \hbar } \left (\beta  \Gamma\right )^{1/2} \: \left ( \frac{ \beta {\cal{J}}} { \pi \alpha _S N }  \right )^{1/2}.
  \label{kapo}
 \enea
 This contribution to the thermal conductance  is $\sim {\cal{O}}\left (\left [ \frac{ \beta {\cal{J}}} { \pi \alpha _S N }  \right ]^{1/2}\right )$ and is reported in the sketch of Fig.\ref{tritoc} as $\sim \sqrt{T}$.  
   The definition of  $ \tilde{a}_\ell ^2 \sim  \tilde{a}^2 \: T_0 / T $, in  Eq.(\ref{tdisst}), emerged  in a previous study \cite{noi}, by including scaling to finite  $ \beta {\cal{J}}/ N $.  Both  $T_{coh}$ and  $T_{0}$ are ${\cal{O}} \left ( \frac{ \beta {\cal{J}}} { N }\right )$ because  they are  parameter scales   of marginal perturbations in the large $N$, large ${\cal{J}}$ limit.  It follows that the contribution to $\kappa$ given by Eq.(\ref{kapo}) is subdominant  with  respect to the  constant one coming from the IR limit given by  Eq.(\ref{kppa}).

\section{ $3-d$ quantum Liquid  in interaction with the SYK system in the temperature range $( T_{coh}, T_0)$   }
In this Section we consider a $3-d$ quantum Liquid (qL), which  would be a single band FL if  in isolation,  and we consider  its interaction with our extended  $2+1-d$ SYK model. 
In Section II  and III we have recognized that there is a temperature threshold $ T_{coh}$ given by Eq.(\ref{tcoh}),  below which a   $3-d$ qL  in interaction with the SYK system  acquires coherence and a threshold temperature $T_0$ given by Eq.(\ref{tdisst}), at which  the Q-excitations  due to the UV corrections, become dominant.  In this Section we discuss how  the fermionic  SYK excitations  and the collective bosonic Q-excitations influence the coherence of the qL.  Our aim is to derive  changes in the lifetime of the quasiparticles of the qL due to the interaction.    We start with  the correction to the selfenergy of the FL  induced by the   energy distribution of the fermionic SYK excitations in the conformal limit (for $T \lesssim T_{coh}$). We will find that the qL becomes a Marginal Fermi Liquid (MFL) (Subsection  A).  Next we derive  the lifetime  reduction  induced by  the interaction with  the bosonic Q-excitations in the temperature range $( T_{coh}, T_0)$ (Subsection B).

\subsection{marginal Fermi liquid (MFL) for  $T \gtrsim T_{coh}$}

In a perturbative approach\cite{chowdhurySenthil}, let us assume a local interaction strength ${\cal{J}}$ in the qL. 
 The quasiparticles of the low energy FL have a  residue $Z$, such that  $Z\nu_0 \sim {\cal{J}}^{-1}$.
   The  single particle energy is  $\epsilon_{\vec{k}}= \tilde{v}_F k $  in the continuum limit, in the  vicinity of the FS,  with a renormalized physical velocity $ v_F^* = Z \tilde{v}_F $.
 The isotropic self-energy arising from the interaction, for $\vec{k}$ on the Fermi surface, is: 
 \bea
   \Sigma (k_F, i\omega )\hspace{6cm}  \nonumber \\
   = {\cal{J}} \sum _{\vec{q}} \int \frac{d\Omega}{2\pi} \:  G( \epsilon_{\vec{k}_F+\vec{q}}- \epsilon _{\vec{k}_F} , i\omega + i\: \Omega ) \: \Pi ( q,i\: \Omega )\nonumber\\
   \approx {\cal{J}}\sum _{\vec{q}} \int \frac{d\Omega}{2\pi} \: \frac{ 1}{i Z^{-1} (\omega + \Omega ) - \tilde{v} _F q \cos\theta }\: \Pi ( q,i\: \Omega ), \label{sigo}
   \enea
   where $ \Pi (q, i \: \Omega ) $ is the polarization function 
    \bea 
      \Pi (q, i \: \Omega ) = \sum _p  \sum _{\omega _n} G( \epsilon_{ p},i \omega_n )\: G( \epsilon_{p+q}, i\: \omega_n + i \: \Omega_m). \nonumber
\enea
 In Eq.(\ref{sigo}), $\theta $ is the angle between $\vec{q}$ and $\vec{k}= \vec{k}_F$ and, for $|q| << k_F$, we have approximated $ \epsilon_{\vec{k}_F-\vec{q}}- \epsilon _{\vec{k}_F} \approx  \tilde{v} _F q \cos\theta$. 

 In the range of frequencies $ \Omega \sim  T_{coh} = W^2/ {\cal{J}}$, where $W$ is the bandwidth  ($k_B = 1$ here), the contribution to the polarization, coming from the virtual  FL  quasiparticles  gives the well known result:
 \bea
\Im m \{  \Sigma^{(1)} (k_F, \omega )\}\} \approx   \alpha \: Z^{-2} \: sign[\omega ] \: \omega ^2  
\enea
($\alpha $ is a parameter of order one).
 The contribution coming from interaction with the extended SYK system, $ \Pi^2 ( i \: \Omega ) $, is evaluated from the single particle Green's function of the SYK model, in the conformally symmetric limit, which is  $q-$ independent and  reported in Eq.(\ref{galt}),
\bea
\Pi ^2 (i\: \Omega ) = - \frac{1}{{\cal{J}}}\!\! \left \{ \! \int _{-{\cal{J}}}^{- T_{coh}} \!\!\!\! + \! \int _{T_{coh}}^{{\cal{J}}} \right \} d\omega\: \frac{ [sign(\omega+ \Omega)\: sign(\omega)]}{\sqrt{|\omega |} \sqrt{|\omega+ \Omega  |}}\nonumber\\
 \approx  - \frac{4}{ {\cal{J}}}\: \ln \left (\frac{ {\cal{J}}}{max[ \Omega,T_{coh} ,T]} \right )\to - \frac{8}{ {\cal{J}}}\: \ln \left (\frac{ {{W}}}{T_{coh}} \right ). \hspace{1cm}
\enea
 The final result is: 
  \bea
\Im m\{ \Sigma ^{(2)}(\omega )\}\approx - \: \frac{  \epsilon _F}{ 2 T_{coh}}\: |\omega| \:\ln \left ( \frac{{\cal{J}}}{T_{coh}}\right ) \: sign (\omega ). 
\label{rel}
\enea 

For $T> >T_{coh}$ we should put $ \ln {{\cal{J}}}/{T_{coh}} \to \ln({\cal{J}}/ T )$ in Eq.(\ref{rel}). 
 $\Sigma^{(1,2)} (k_F, \omega ) $ changes sign at $\omega =0$ when the quasi-particle becomes a quasi-hole. Putting both terms together, the quasiparticle relaxation rate is: 
\bea
 \frac{1}{\tau} \sim -Z \: \left | \Im m  \Sigma (k_F, \omega ) \right |\nonumber\\
 = |\omega| \frac{Z \epsilon _F}{ k_BT_{coh}} \ln \left ( \frac{{\cal{J}}}{ T_{coh}}\right )\!+\! \frac{ \alpha }{Z} \:\nu_0\:  |\omega |^2,
 \label{tsca}
 \enea
which shows that, to the lowest approximation, the perturbed FL is a Marginal Fermi Liquid.  The lowest lying collective excitations of the MFL can be also derived  in the present perturbative frame \cite{noi}.  An  hydrodynamic collective excitation, the would-be acoustic plasmon, is  rather well defined. The acoustic plasmon is on the verge to emerge as a bound state at low energies, splitted off the p-h continuum. 
 Its dispersion tends to the boundary of the p-h continuum, where the imaginary part vanishes. 
 
 In Section II.C we have derived a thermal conductance  $\kappa$  in the conformal limit, which is independent of $T$. This $T-$independence   appears to be consistent with the present  result. 
  Semiclassically,  the left hand side of Eq.(\ref{vir13}) can be approximated as ($\omega \sim 0$):
     \bea
       \met \int d^2r \int dt \int dt'\: \left  \langle  \left \{ J^{{\cal{E}}}(r,t) ,J^{{\cal{E}}}(r,t')  \right \}  \right \rangle \nonumber\\
       \approx   \overline{n}_{MFL}\: \frac{L}{2\pi} \tilde{v}_F^2 \:  \langle \left ( \delta {\cal{E}}\right )^2 \rangle\:  \tau _{MFL} 
       \label{sem}
       \enea
        The saddle point contribution to energy of the SYK model of Eq.(\ref{correneg}) provides us with the energy fluctuation,  given by:
    \bea
  \langle \left ( \delta {\cal{E}} \right )^2 \rangle = \partial _\beta^2 \ln Z = \frac{c}{\beta^3}.
  \label{fluttE}
  \enea
  By plugging  Eq.(\ref{sem}) in  Eq.(\ref{vir13}) with Eq.(\ref{fluttE}), we get,  in the limit $\hbar \omega /k_BT <<1$, 
       \bea
\kappa  \approx  k_B \: \frac{L}{2\pi}  \overline{n}_{MFL}\:  \tilde{v}_F^2\: \frac{1}{(k_B T)^2}  \frac{ 4\pi^2 \alpha _S N}{\beta^3{\cal{J}}}\: \tau _{MFL} .
\label{ktMFL}
 \enea
The MFL  lifetime of Eq.(\ref{tsca})  is  $\tau _{MFL}\sim T^{-1}$,  so that  
  $T-$ power-counting  
gives the  $\kappa $ contribution  of Eq.(\ref{ktMFL}) independent of $T$, at this perturbative level. However this contribution   to the thermal conductivity is again subdominant  as $   \sim {\cal{O}} \left ([\beta {\cal{J}}/N ]^0\right )$, while the contribution coming from the IR limit of Eq.(\ref{condur}) is  ${\cal{O}} \left (\beta {\cal{J}}/N \right )$ .

 \subsection{Influence of the Q-excitations on the lifetime of the MFL ( $T \gtrsim T_{coh})$}

The Q-excitations introduced in Section III.B are an additional scattering mechanism which contributes to the reduction of  the lifetime of the quasiparticles of the qL. 

According to the source term of  Eq.(\ref{essource}),  $J^{\cal{E}}-J^{\cal{E}}$  fluctuations  couple to the change of the chemical  potential,  as can be read off   Eq.(\ref{muo}).  Changes in the chemical potential perturb the qL. Our aim in this Subsection is to derive the effect on the  lifetime of the quasiparticles  of the qL  due to this kind of interaction. The thermal average of  Eq.(\ref{muo}) can be newly defined as: 
\bea
{G}_{+}^{LQ}(q, i\: \omega ) =  \frac{h}{\rho_0}  \frac{\int dk' \int d\omega ' \: J^{\cal{E}} _{-k' ,-\omega'}J^{\cal{E}}_{k' ,\omega'} }{(k_BT)^2} \: \frac{-i\: \omega  }{ \tilde{\cal{D}} \: q^2 -i \omega}.\hspace{0.2cm}
\label{sourc}
\enea
 In real time ${G}_{+}^{LQ}$  is a response function  $  G_+(t )  = \Theta (t)  G^(t) $ and  has the dimensions of   $[{\cal{E}}]$. 
 
 Here we face a problem in the construction of the interaction matrix element between the SYK sheet system and the $3-d$ qL.  The transferred momentum $q$ is  a $3-d$ vector  in the liquid, but it is only defined in two dimensions  in the  $ J^{\cal{E}} -J^{\cal{E}} $ term. We overcame the same  difficulty in Eq.s(\ref{vir13},\ref{res}), by  considering the convolution (with $k-$ integration) as  in Eq.(\ref{cocori}) of Subsection III.B, which corresponds to averaging in space, but by assuming that   the fluctuation correlations are  localized within a volume $\sim \tilde{a}_\ell^2 $.  We eventually  get 
    \bea
    G_+^{LQ}(q, i\: \omega ) = 
    2\pi   \frac{h}{ \rho_0v^2 \tau _0^2 } \:\frac{ \tilde{a}_\ell^2 }{ \tilde{a}^2}\:  \frac{\Gamma}{\hbar}\:  \frac{-i\omega }{-i\: \omega +  \hbar \tilde{\cal{D}} q^2 }.
        \label{simio}
    \enea
   Assuming  in  the     $T-$power counting  $ \rho _0 v^2 \tau _0^2  \sim {\cal{O}} \left ([T ]^0\right )$, with $\hbar/\tau_0 \sim  \Gamma \sim T$,  the prefactor of  the energy $G_+^{LQ}$, provides a squared coupling prefactor  $|\tilde{g}|^2 \sim  T^0 $  while,  in  the case  of  the e-ph interaction in a metal,   $ |\tilde{g}|^2 \sim T $. 
  
Eq.(\ref{simio}) can  play the role of an  interaction energy  matrix element  for   the scattering between Q-excitations and a MFL quasiparticle. We   plug it   in the Fermi Golden Rule and  derive the thermalization  rate\cite{schofield}($\omega $ has dimension $ [{\cal{E}}]$  from now on):
\bea
\frac{1}{\tau_\kappa}  =\frac{2\pi}{\hbar } \!\! \left (\frac{L}{2\pi}\right )^d \!\!\int _0^\epsilon  \!\!\nu_0\: \omega \: d\omega \int _{\omega / \hbar v_F} ^{2k_F} \!\! \frac{ q^{d-1}\: dq \left [{G}_{+}^{LQ}(q,i\: \omega ) \right ]^2}{( \hbar v_F q)^2}.\nonumber\\
\label{teps}
\enea
 
 $\nu_0$ is the density of states at $\epsilon _F$  for  a $3-d$ volume.
The integration over the direction of the momentum has already been performed and gives the factor of $( \hbar v_F q)^2$  in the denominator, reflecting the increased
time available for small deflections.  As a minimum momentum must be transferred to give a change in energy of $\omega $,  $\omega $  appears also  in the
lower limit of the momentum integral.   The integral over $\omega $ is simply the number of possible hole excitations that can be created in the   qL close to  the FS. If  $G_+^{LQ} (q,\omega  )$   is independent of $q$ and $\omega $, the $q$ integral is not sensitive to the value of the lower limit and is independent of $\omega $. The  subsequent integration over $\omega $ recovers the usual  $3-d$ result,  $ {\tau_\epsilon}^{-1} \sim \epsilon ^2$, which is  expected for a FL.

Inserting $G_+^{LQ}$  of Eq.(\ref{simio}) in  Eq.(\ref{teps}), we have

  \begin{widetext}
      
     \bea
\frac{1}{\tau_\kappa}  =\frac{2\pi}{\hbar } \: \left (\frac{L}{2\pi }\right )^2 \int _0^\epsilon \nu_F \omega\: d\omega      \left ( \frac{\tilde{a}_\ell}{\tilde{a}} \right )^4  \left ( \frac{h}{ \rho_0 v ^2\tau _0^2}\right )^2 \left (  \omega \frac{\Gamma}{\hbar} \right )^2\: 
\frac{L}{2\pi} \int _{\omega / \hbar v_F } ^{2k_F} \frac{dq}{(\hbar v_F)^2}  \left | \frac{ 1}{\hbar \tilde{\cal{D}} \: q^2 -i \omega}\right |^2,
\nonumber
\enea
\end{widetext}
 which gives 
        \bea
\frac{1}{\tau_\kappa}  =\frac{2\pi}{\hbar } \!\!  \left (\frac{L}{2\pi \hbar v_F}\right )^3  \!\!\!\int _0^\epsilon \nu_F \omega^2 d\omega \: \left ( \frac{\tilde{a}_\ell}{\tilde{a}} \right )^4  \!\!  \!\!  \left ( \frac{h}{ \rho_0 v ^2\tau _0^2 }\right )^2  \!\!  \!\!  \left (  \frac{\Gamma}{\hbar} \right )^2 \!\!  \!\! . 
\nonumber\\
\label{tki}
\enea
In evaluating the integral,  we have excluded the upper limit,  because it implies very large  transferred $q$ values, which we neglect as  it gives a contribution ${\cal{O}}\left ( \epsilon ^4\right )$ to the result. 
To proceed to $T-$power counting,  we assume that the  volume is factorized into an in-plane contribution  $  L^2 \sim {\tilde{a}_\ell ^2} $  where the diffusion dynamics takes place, times  an out of plane one, represented by the term $ \nu_0 \hbar v_F L / \tilde{a}_\ell ^2 $.  Assuming again  $ \rho_0 v ^2\tau _0^2\sim T^0 $, we get:
 \bea
\frac{1}{\tau_\kappa}   \propto  T^2 \:\left ( \frac{ T_0}{T} \right )^3 \epsilon ^3 \sim T^2.
\label{tpok}
\enea
This $T-$dependence  cannot be distinguished from the FL one, although it arises from a very different origin,  that  is  from  the scattering of  MFL  single particle excitations  by the Q-excitations. Just  for comparison the electron -phonon scattering rate, is\cite{ashcroft}: $ {\tau_{e-ph} } ^{-1}  \sim T^3 $ for $ T << \theta _D$.
 
If we use Eq.(\ref{ktMFL}) as an approximate expression for the thermal conductance, in which we insert the result of Eq.(\ref{tki}),  we get  $ \kappa  \to  1/T$ as in the FL case. In fact, 
in   the FL case,  $\kappa = C \: v_F \ell = C \: v_F^2 \tau_0$. As  $ \tau_0 \sim T^{-2} $ and  $ C \sim T$, then  $\kappa  \sim  T^{-1}  $. However, as $\tilde{a}_\ell^4$ appears in Eq.(\ref{tki}),  this $\kappa$ contribution    $\sim {\cal{O}}\left ([ N/ \beta {\cal{J}}]^3\right )$ and it is irrelevant for large $\beta {\cal{J}}$. 




 On the other hand, resistivity  is ruled by a  relaxation time  in which   the forward scattering  is  subtracted from the response appearing in Eq.(\ref{simio}). For very small $\omega$,  both  integrals are  convergent and we have: 

    \bea
\frac{L}{2\pi} \int _{\omega / \hbar v_F } ^{2k_F} \frac{dq}{(\hbar v_F)^2}  \left | \frac{ -i\omega }{\hbar \tilde{\cal{D}} \: q^2 -i \omega} -1 \right |^2
=   \frac{1}{2\pi }\:  \frac{2k_F L}{(\hbar v_F)^2} ,  \: \nonumber
\enea
    where we have excluded the bottom  limit of integration. Hence: 
   
    \bea
\frac{1}{\tau_\sigma}  =\frac{1}{\hbar } \: \frac{\tilde{a}^2_\ell  L}{( \hbar v_F)^3} \!\!  \int _0^\epsilon \nu_F \omega d\omega    \left ( \frac{\tilde{a}_\ell}{\tilde{a}} \right )^4  \!\!  \left ( \frac{1}{ \rho_0 v ^2\tau _0^2 }\right )^2   \!\!  2\: \Gamma ^2\hbar v_F k_F.\nonumber\\
\label{tsi}
\enea

Here the $T-$power counting provides   $\frac{1}{\tau_\sigma}   \propto  T^2 \:\left ( \frac{ T_0}{T} \right )^3 \epsilon ^2$ $\sim  T$  which points to the unbound dependence   $ \propto T$  of the resistivity, once more.  This result  should be compared with the fifth $T-$power  arising from the electron-phonon scattering and  shows that,  although   the constraints on momentum conservation  has been relaxed, the localized  nature of the Q-excitations does not   increase resistance  with increasing temperature any further.  This can also be seen from the presence of $\tilde{a}^6_\ell$ in Eq.(\ref{tsi}) which implies that a contribution to the resistivity would drop at increasing temperature as $  {\cal{O}}\left ([ N/ \beta {\cal{J}}]^3\right )$, again,  and soon become irrelevant.  

 \section{Summary and Conclusions}

 The SYK model, borrowed from  gravity theories,  is  solvable at large flavour number $N$  and is intrinsecally NFL.  Non vanishing  zero temperature entropy is the signature of a crucial role of quantum fluctuations in the build up of  the ground  state of the system. 
The model is originally in $0+1-d$ dimensions  and its fermionic fields are Majorana neutral fermions. This feature, if taken seriously,  already implies that  the  charge and spin degrees of freedom have been dropped out of the model\cite{lantagne}, when describing  the electronic carriers. Indeed Majorana Fermions are neutral and  interaction  ${\cal{J}}$ grows to  infinity but  is fully local  and momentum is not conserved. As transport on the c-axis could be attributed to tunnelling, we propose in this work an extension of the SYK model to two space dimensions and  include  its interaction with a three dimensional quantum Liquid (qL).   

 Our focus is on the symmetry breaking  of an approximate   conformal symmetry of the SYK model in the IR, strong coupling limit. For temperature  in the neighborhood of $ T_{coh}$, the  symmetry breaking  generates a "quasi-order parameter" $G_c ( \tau )$  and "single" particle  neutral fermionic excitations which  are responsible for transforming the qL into a marginal Fermi Liquid. The  linear  dependence of the resistivity  on  $T$  and the independence of $T$ of the thermal conductivity, derived with  the linear response, follow. However, the spontaneous breaking of the conformal symmetry is in reality  explicitely produced by UV  corrections. The UV corrections are local in time and they are found to be  extended, but rather localized   in space,  when the SYK model is extended to higher space dimensions.  They give rise to  bosonic collective modes which we nickname as Q-excitations,  in interaction with the qL.  The Q-excitations are diffusive modes in the lattice. We assume that they have  lifetime $ \tau_0 \propto \hbar \beta$ and  we infer by  scaling\cite{noi} that they have mean square space extension $ \tilde{a}_\ell ^2  \sim T_0/T $.  In this work we have derived the contribution of these excitations to the  lifetime of the qL quasiparticles. The trends  in the temperature dependence of the resistivity $\rho$ and of the  thermal conductivity $\kappa$  derived in our approach  are reported in   Figure \ref{tritoc} for  the  temperature  window $ T_{coh} \lesssim T \lesssim  T_0$, in which there is a crossover to the incoherent regime. For $T > T_0$  the SYK model  saturates a bound on  chaos\cite{maldacenaShenker}  and no perturbative approach is any more feasible.   
 
   To sum up,  we have studied the effect of interactions  between  the SYK model extended  to $2-d$ and the quasiparticles of a $3-d$ quantum Liquid in which the SYK system is embedded.  The  $2-d$ SYK model can be thought of as  a collection of disordered dots in a continuum limit,  in which strong disorder restores space invariance due to self-averaging. In Section II, we have introduced the Hamiltonian of the SYK model extended to the  lattice. The temperature threshold, $T_{coh}$, below which the hopping in the lattice establishes coherence in the system  has been  defined  in Section II.A and the complex fermion occupation number $n_k$ close to the Fermi energy has been   derived  in the conformal symmetry limit, by  including the lattice perturbation.  We show that these complex fermion single particle  excitations which remind of FL quasiparticles are ill defined at the Fermi Surface, confirming the NFL nature of the extended version of the SYK model. This fact justifies our choice of concentrating on the neutral excitations of the model,  which are the  fermionic excitations originating from the  IR saddle point  strong coupling limit and the collective diffusive Q-excitations arising from UV corrections to the IR theory. The generating functional of the  soft mode correlations, derived  in the gaussian approximation\cite{noi}, is reported  in Section II.B.  An hydrodynamics  of energy, instead of particles is introduced  in this Section. It is crucial to keep in mind that the underlying physics of the  incoherent SYK system  implies that the $2-d$ system acts as a sink for particles  and momentum, so that  the usual particle hydrodynamics is not justified. On the contrary, the energy current density has a  diffusive dynamics which does not require  momentum and particle conservation. The source term to be added to the action is derived in this way and is used in  Section II.C,  to obtain  the thermal conductivity $\kappa$ and the resistivity $\rho $ of the model in the conformal symmetry limit\cite{song}. These transport coefficients are proportional  the temperature scale $T_{coh}$ which is 
   $\sim {\cal{O}}\left ( \beta {\cal{J}}/N\right )$ in the large $N $, large ${\cal{J}}$ limit. 
   
    The role of the  UV corrections in giving dynamics to the  collective  bosonic excitations, the  Q-excitations, is presented  in Section III.  The Q-excitations  can be shown to be  diffusive in  $2-d$\cite{noi} and to produce  energy current density fluctuations  which become dominant for $T \sim T_0$. The response of the chemical potential  to these fluctuations  is derived in Section III.A.  A Lagrangian approach to the dynamics of the  Q-excitations, which includes diffusivity in an approximate way, and their quantization is presented in Section III.B.  The $T$ dependence of their contribution to transport   for $T\sim T_0$ follows from the choice of  their lifetime  $ \tau_0 \sim \hbar \beta$, which implies  full thermalization of the SYK system. It turns out that this contribution to thermal conductivity is subdominant  
    ($\sim {\cal{O}}\left ([ \beta {\cal{J}}/N]^{1/2}\right )$) with respect to the one of Section II.C.
    
     Section IV  is dedicated to the  $3-d$  qL and to  its interaction both with the single particle  fermionic excitations\cite{chowdhurySenthil}  and with the Q-excitations  introduced in Section III. The lifetime of the quasiparticles of the qL  is derived in  Section IV.A.   The  interaction  with the  conformally symmetric SYK  system   is dealt with perturbatively, and makes the liquid a MFL. The  contribution  of the MFL to the  thermal conductivity is derived  in a semiclassical approximation and is found to be consistent with the one derived in Section II.C.  In Section IV.B  the influence of the Q-excitations on the lifetime of the MFL is derived  by using  the Fermi Golden Rule. These scattering rates allow to estimate  the resistivity and thermal conductivity  for  $T\sim T_0$.  The estimates  of  the temperature dependence  of  $\kappa$ and $\rho $  are  visualized in  the sketch of Figure \ref{tritoc}. 
     
     The model confirms, also beyond the IR limit, the temperature dependence of the resistivity  that is expected in the measurements,  if the model does indeed entail features of the real "strange"  NFL metals. All additional contributions to the thermal conductivity estimated here are subdominant with respect to  the IR limit  contribution  (Eq.(\ref{condur})), which is of $ {\cal{O}} \left (\beta {\cal{J}}/N\right )$ in the large $N$, strong interaction ${\cal{J}}$ limit.    The pivotal quantity of our hydrodynamic approach to the collective Q-excitations is their mean square diffusion area  $ \tilde{a}_\ell^2  \sim T_0/T$. In the scaling to the strong  interaction limit, both $T_0$ and  the diffusion coefficient $\tilde{D}_Q$ are temperature independent. The incipient localization with increasing temperature locates our result straightforwardly out of the Mott-Ioffe-Regel   resistivity  limit\cite{hussey}.     However, for  $T > T_0$ the $0+1-d$ SYK model  enters into a scrambled phase which is fully chaotic.  The analysis of this phase in the extended model  is out of  scope of this work.

 \section*{Acknowledgements}
 The author acknowledges useful discussions   with F.noi,  Elisa Ercolessi and Francesco Tafuri and financial support of Universit\`{a} "Federico II" di Napoli, project "PlasmJac" E62F17000290005 .


 \appendix
 
 \section { Relativistic approach to transport}

In a particle system at equilibrium, the free energy per particle  is  $ \frac{F}{N } =  \frac{U}{N }  - \frac{TS}{N } $, where  $U/N= \epsilon$ is the internal energy per particle and $s= S/N$ is the entropy per particle. Adding and subtracting ${p}/n$ where ${p}$ is the pressure and $n$ is the particle density, equilibrium implies: 
   \bea
     \:\:\:\:   dF = 0  \:\: \to  \:\: d\left ( \frac{w}{n} \right ) =  T \: d  \left (\frac{s }{n}\right ) +\frac{d{p}}{n}. 
     \label{eqstat}
     \enea     
     Here $w= \epsilon +{p} $ is the enthalpy and small letters are thermodynamic quantities per unit volume. 
     
     Besides, in terms of the Gibbs potential, $  G= \mu N  = F+pV  $  we write 
     \bea
     \frac{w}{n } - \frac{T\:s}{n } = \mu,  
     \label{popo} 
     \enea
where $\mu $ is the chemical potential.  

Here we consider  heat flow, even in the absence of matter convection, starting from  $  T_i ^k $, the 
energy-momentum tensor:  
\bea
T_i ^k= w \: u_i u^k + p \: g_i^k + \tau _i^k , 
\label{tens}
\enea
where $g_i^k $ is the metric and  $ \tau _i^k $ is the stress tensor.  In the proper frame, the velocity $u^\alpha =0 $ and $u^0 =1$ (italic indices $ i=0,1,2,3$, while  greek indices $ \alpha =1,2,3 $). Separating the center of mass flux from a small fluctuating contribution,   we have, as in  Eq.(\ref{cora}):   
 \bea
\vec{J} = n \vec{u}  + \vec{\nu}  \to  T_\alpha^0, \nonumber\\
 \vec{J}^{{\cal{E}}} = \frac{( \epsilon +p )}{n} \: \left (\vec J -\vec \nu \right )  = T_0^\alpha \equiv  w \: u^\alpha, \nonumber
 \enea
  where $ \vec{J} $ is the particle current density while $ \vec{J}^{{\cal{E}}} $ is the energy current density.  
 In the proper frame,  the vector particle flux density  $ n u^k + \nu^k $ implies  $  \nu^k  u_k =0 $, because its  0 component must be equal to the particle number density $-n$ :   $ n \: u^k  u_k +   \nu^k  u_k  =-n $, by definition. The  continuity equation reads 
 \bea  
   \frac{\partial }{\partial x^k}  \left [ n\: u^k +  \nu^k \right ] =0 .
   \label{conto}
   \enea
    $  T_i ^k $, includes  dissipative processes (viscosity and thermal diffusion).
 However, in the absence of external forces, conservation requires  $
 \frac{\partial}{\partial x^k} \: T_i ^k =0 $, or \cite{landauHydro}
 \bea
   0 = u^i \frac{\partial}{\partial x^k} \: T_i ^k =  u^i \frac{\partial p}{\partial x^k} \: g _i ^k 
   + u^i u_i  \: n \: u^k \frac{\partial w/n}{\partial x^k} \nonumber\\
   + u^i u_i  \: \frac{w}{n}  \:\frac{\partial n u^k}{\partial x^k} +
   u^i w  \frac{\partial u_i}{\partial x^k} u_k   +     u^i  \frac{\partial }{\partial x^k} \tau_i^k .
   \label{serv}
   \enea
   Here $ u^i u_i  =-1 $,  so that $ u^i  \frac{\partial u_i}{\partial x^k}  =0 $ and  the fourth term vanishes.  In the third term we substitute $ \frac{\partial n\: u^k}{\partial x^k}  \to - \frac{\partial \nu^k}{\partial x^k} $, according to Eq.(\ref{conto}).     
   Next, from  Eq.(\ref{eqstat}),  we have: 
    \bea
 u^i  \left [  \frac{\partial p}{\partial x^i}  -    n\:  \frac{\partial w/n}{\partial x^i}  \right ]  = - T \: n\: u^i \frac{\partial s/n}{\partial x^i}   
  \label{eqstat1}
\enea
and  Eq.(\ref{serv}) becomes: 
   \bea
   0= -T  n\:  u^i   \frac{\partial }{\partial x^i} \left (\frac{s }{n}\right )+ \frac{w}{n}  \:\frac{\partial \nu^k}{\partial x^k} +   u^i  \frac{\partial }{\partial x^k} \tau_i^k,\nonumber
   \enea 
   or, using Eq.(\ref{popo}), 
    \bea
     0= -T   \frac{\partial }{\partial x^i} \left (s \: u^i \right )+ \mu \:\frac{\partial \nu^k}{\partial x^k} +   u^i  \frac{\partial }{\partial x^k} \tau_i^k .
    \label{inter}
    \enea
    The dissipative components of the stress-energy tensor are orthogonal to $\vec{u}$:  $ u^i \tau _i^k =0 $  and, from   Eq.(\ref{inter}), we get:
    \bea
T\left \{ \frac{\partial}{\partial x^i}\!\left ( s\:  u^i  \! - \!\frac{\mu}{T}  \nu ^i \right ) \!\right \} = - T \nu ^i  \! \frac{\partial }{\partial x^i} \left (\frac{\mu}{T} \right ) \!- \! \tau _i ^k \frac{\partial u^i}{\partial x^k}.
\label{cruk}
\enea
Eq.(\ref{cruk})  is the crucial equation  in the absence of external forces, originating from  $ u^i \frac{\partial}{\partial x^k} \: T_i ^k =0 $.
On the l.h.s. there is the variation of the entropy flux density which has to be positive.  This qualifies, to lowest order, the expressions for the dissipative components of the stress-energy tensor  $\tau _i^k$ and the current density $\nu^k$.
 The component of the current are:
  \bea
\nu _i = -\kappa \left ( \frac{n T }{w} \right )^2 \:\left [ \frac{\partial}{\partial x^i} \: \left (\frac{\mu}{T} \right ) + u_i u^k \frac{\partial}{\partial x^k} \: \left ( \frac{\mu}{T} \right ) \right ]. 
\enea
 $\kappa $  is the heat current response  to $-\nabla T $ in the absence of a driving particle  current.  The choice of the prefactor is in agreement with the Wiedemann-Franz  law $ \frac{\kappa}{\sigma T } = \frac{3}{2} \left ( \frac{k_B}{e} \right )^2 
$.
Thermal conduction with no particle flux   implies   vanishing of the second term.  We have: 
\bea
  T_\alpha ^0 =  w  u^0 u_\alpha = - \frac{w}{n} \: \nu _\alpha = \frac{\kappa n T^2}{w} \frac{\partial }{\partial x^\alpha}  \left ( \frac{\mu }{T} \right )\nonumber\\
 = -\kappa \left [ \frac{\partial  T}{\partial x^\alpha}  -  \frac{T}{w} \frac{\partial  }{\partial x^\alpha} p \right ].
 \label{moti} 
  \enea
       The last equality arises from the following manipulations:
  \bea
 d \left ( \frac{\mu }{T} \right )
= -\frac{ 1}{T^2 } \frac{w }{n}  \: dT  - d\left ( \frac{ s}{n } \right ) + \frac{1 }{T} \: d \left [  \frac{T\: s }{n} +  \mu  \right ]\nonumber\\
=    -\frac{ 1}{T^2 } \frac{w }{n}  \: dT  +\frac{1}{n\: T } \: d p,\nonumber
 \enea
 where Eq.(\ref{eqstat}) has been used. 

  Eq.(\ref{moti} is the energy flux, which also includes a $ \nabla p $ term that is absent  in  the non relativistic result\cite{landauHydro}. 
  
    From linear response, the energy density current $ J^{\cal{E}} $ can be written as\cite{tagliacozzoCampagnano}: 
   \bea
 J^{\cal{E}} 
 = w\: J +  \kappa\:  \nabla T +  \frac{w}{n}\frac{\sigma_{\ell\ell} }{e^2}\:  \left [ -\nabla p \right ] \nonumber\\
  +\frac{ w}{n} \: \hat {\ell}  \frac{\sigma _{\ell i}}{e^2}  \epsilon _{ij3}   J_j  \frac{1}{c^2}\oint d\vec{\ell} \cdot \vec{J},
  \label{cure}
 \enea 
where $\sigma _{ij}$ is the conductivity tensor with  $ \hat{i}, \hat{j}, \hat{\ell}$  unitary vector in real space.   We have added the last term, which  is the Magnus force. The Magnus force is present  in case there is a current  circulation  $\oint d\vec{\ell} \cdot \vec{J}  \neq 0$  in the flow. In the case of an electromagnetic current it would take the form  $ w\: e J \times \frac{B}{n\:e\: c} \frac{\sigma _H}{e^2} $, when induced by a magnetic field $B$, with  $\sigma _H$ being  the Hall conductance\cite{hartnollKovtun}. The dimension of  $[J^{\cal{E}} ]$ is $ \left [ \frac{{\cal{E}}}{\ell^{d-1} t} \right ]$.
 
 We can define the compressibility  $\chi$ and the pair of diffusion constants  for the Q-excitations  $ \tilde{D}_Q$  and for the particles  $ \tilde{D}_e$, respectively,  according to: 
  \bea
  \chi =- \frac{1}{\cal{V}} \frac{ \partial {\cal{V}}}{\partial P},
   \:\:\:\:  \tilde{D}_Q  = \frac{ \kappa}{\rho_0 C} ,    \:\:\:\:  \tilde{D}_e  = \frac{ \sigma }{\chi }, \label{ded}
   \enea
   (space isotropy is assumed here for simplicity). The diffusivity $\tilde D_Q$ is related to the thermal capacitance $C$ and  $\rho_0$ is a particle density  associated to the Q-excitations which does not concide with the particle density $n_e$ of the carriers and is defined after Eq.(\ref{caco}).  A generalized diffusion constant $ \tilde{D}$ can be defined, according to: 
   \bea
   \frac{1}{n}\:J^{\cal{E}} 
 =   \tilde{D}_Q  \:C \nabla T +  \frac{w}{n^2}\frac{\sigma }{ \chi e^2}\:\left [  \frac{1}{\cal{V}} \frac{\partial  {\cal{V}}}{\partial T } \right ] \nabla  T
\equiv \tilde{\cal{D}} \: C \:  \nabla  T.\nonumber\\
\label{mott}
 \enea

\bibliography{./biblioscotto2}
\end{document}